\documentclass[fleqn,english,usenatbib]{mnras}

\usepackage{newtxtext,newtxmath}

\usepackage[T1]{fontenc}
\usepackage{ae,aecompl}

\usepackage[T1]{fontenc}
\usepackage[latin9]{inputenc}
\usepackage{color}
\usepackage{babel}
\usepackage{units}
\usepackage{textcomp}
\usepackage{amsmath}
\usepackage{amssymb}
\usepackage{graphicx}
\usepackage{breakurl}
\usepackage{savesym}
\usepackage{amsmath}
\usepackage{tabu}
\usepackage{longtable}
\usepackage{supertabular}
\usepackage{nicefrac}
\usepackage[usenames,dvipsnames,svgnames,table]{xcolor}
\usepackage{pifont}
\usepackage[tableposition=top]{caption}

\providecommand{\tabularnewline}{\\}
\savesymbol{iint}
\savesymbol{iiint}
\savesymbol{iiiint}
\let\oldAA\AA
\renewcommand{\AA}{\text{\normalfont\oldAA}}

\title[Primordial helium abundance determination using sulphur as metallicity tracer]{Primordial helium abundance determination using sulphur as metallicity tracer}

\author[V. Fern\'andez, E. Terlevich, A. I. D\'\i az, R. Terlevich, F. F. Rosales-Ortega]{Vital Fern\'andez$^{1}$\thanks{E-mail:
vi.fernandez@inaoep.mx (Vital Fern\'andez)}, Elena Terlevich$^{1}$, Angeles I. D\'\i az$^{2, 3, 4}$, 
Roberto Terlevich$^{1, 5}$, \newauthor F. F. Rosales-Ortega$^{1}$\\
$^{1}$Instituto Nacional de Astrof\'\i sica, \'Optica y Electr\'onica, Luis E. Erro 1, 72840 Tonantzintla, Puebla, Mexico\\
$^{2}$Departamento de F\'isica Te\'orica, Universidad Aut\'onoma de Madrid, E-28049 Madrid, Spain\\
$^{3}$Centro de Investigaci\'on Avanzada en F\'isica Fundamental CIAFF-UAM \\
$^{4}$Astro-UAM, UAM, Unidad Asociada CSIC \\
$^{5}$ Institute of Astronomy, University of Cambridge, Cambridge, UK}

\date{Accepted XXX. Received YYY; in original form ZZZ}

\pubyear{2018}

\begin{document}
\label{firstpage}
\pagerange{\pageref{firstpage}--\pageref{lastpage}}
\maketitle

\begin{abstract}
The primordial helium abundance  $Y_P$  is calculated using sulphur as
metallicity tracer in the classical methodology (with $Y_P$ as an extrapolation of $Y$ to zero metals). The calculated
value, $Y_{P,\,S}=0.244\pm0.006$, is in good agreement with
the estimate  from the Planck experiment, as well as, determinations
in the literature using oxygen as the metallicity tracer. The chemical
analysis includes the sustraction of the nebular continuum  and of the
stellar continuum computed from  simple stellar population synthesis grids.
The $S^{+2}$ content is measured from the near infrared $\left[SIII\right]\lambda\lambda9069\AA,9532\AA$
lines, while an $ICF\left(S^{3+}\right)$ is proposed based on the
$Ar^{3+}/Ar^{2+}$ fraction. Finally, we apply a multivariable linear regression
using simultaneously oxygen, nitrogen and sulphur abundances for the same sample
to determine the primordial helium abundance resulting
in $Y_{P-O,\,N,\,S}=0.245\pm0.007$. 
\end{abstract}



\section{Introduction}

The primordial light element abundances $D$, $^{4}He=Y_{P}$, $^{3}He$,
$^{7}Li$ constitute a keystone in  Big Bang cosmology along with
the Cosmic Microwave Background (CMB) and the Universe expansion.
A good agreement between observed abundances, and model parameters
(such as $\eta$, the baryon to photon ratio, the number of neutrino
families or the neutron half-life) enables a detailed definition of
the Big Bang boundary conditions.

Almost twenty five years ago, precise measurements for the $\nicefrac{D}{H}=3.40\pm0.25\times10^{-5}$ by
\cite{burles1998} yielded a deuterium abundance almost a tenth
of those previously reported \citep{songaila1994,carswell1994}. These
observations caused a great commotion, since either higher $Y_{P}$
values, or new primordial nucleosynthesis models or non-standard physics
would be required for consistency. The exploration of these possibilities
also led to a thorough quantification of the systematic errors in
the $Y_{P}$ determination. Overall, all these dedicated studies reached
similar conclusions \citep{olive2004,fukugita2006,izotov2006-a,rosales-ortegaf.f.2006,peimbert2007-a},
providing values consistent with those of other cosmological parameters,
even under the tighter WMAP and more recently Planck experiment limits \citep[see][]{planck-collaboration2015}.

The classical method to determine the primordial He abundance $Y_{P}$,
proposed by \cite{peimbert1974} and \cite{lequeux1979}, involves
the extrapolation of the present helium abundance $Y$ down to zero metallicity. In practice, a
single element, usually $O$, is used as metallicity tracer in ionized
gas. The method is more reliable the lower
the overall metal content of the gas to analyze. This situates 
low metallicity HII regions and local HII galaxies as the best objects
for the job.

HII galaxies are compact, can be easily detected and among them the
lowest metallicy galaxies are found. Their spectra are completely dominated
by the current star formation episode and due to their intense gaseous
emission spectra and weak underlying stellar continua, they can provide
precise measurements of the He emission lines. In fact, in spectra
with $EW\left(H\alpha\right)>200\AA$, the underlying stellar continuum
can hardly be noticed in the optical range, which minimizes this
difficult to quantify source of systematic error in the derived
He abundance.

In these chemically unevolved objects, it is acceptable to assume
a linear relation between the helium content and the oxygen abundance
$\nicefrac{\Delta Y}{\Delta Z}\approx\nicefrac{\Delta Y}{\Delta O}$.
Measuring the helium mass fraction and the O abundance of the gas
for a given galaxy sample, the coefficients of the resulting linear
regression are determined, i.e. $Y_{P}$ is the ordinate extrapolation to zero metals of the regression.
Recently this method to determine  $Y_{P}$ has been used
in \cite{aver2015} and \cite{peimbert2016} yielding
results $Y_{P}=0.2449\pm0.004$ and $Y_{P}=0.2446\pm0.0029$ respectively, consistent with  
Standard Big Bang Nucleosynthesis model predictions and with the value estimated by \citet{planck-collaboration2015}  $Y_{P}=0.24467\pm0.0002$. 
On the other hand, the result by \cite{izotov2014}, $Y_{P}=0.2551\pm0.0022$,
is slightly outside 
the Planck collaboration value\footnote{The Planck experiment provides four $\Lambda$CDM $Y_{P}$ estimations depending on the various data combinations. The value tabulated here represents an upper limit as considered by \cite{peimbert2016}}. The three $Y_{P}$ determinations by \cite{izotov2014,aver2015,peimbert2016} use the latest values of recombination coefficients for the HeI lines
by \cite{porter2012} and the associated corrigendum \cite{porter2013}. 

As already mentioned, most of the research in this topic has used oxygen as the
metallicity tracer in the Y vs. metal regression. Indeed, within the optical range, this element
represents the obvious choice. However, it also presents clear difficulties.
Firstly, while oxygen has the most intense lines in the optical range,
the derivation of the electron temperature necessary for abundance
determinations requires the detection and measurement of the weak
auroral $\left[OIII\right]\lambda4363\AA$ line. This implies the
selection of high excitation, high electron temperature objects for
which a larger collisional excitation contribution to the hydrogen
recombination lines is expected and this effect is challenging to quantify. On the
other hand, due to the required  high excitation, there is a maximum
oxygen abundance of the objects to be used. This, in turn, decreases
the  baseline for the linear regression,  increasing the $\left(\nicefrac{\Delta Y}{\Delta Z}\right)$
gradient uncertainty.

Nitrogen was proposed as an alternative tracer by \cite{pagel1992}
who concluded that despite the nitrogen secondary production mechanism,
a good concordance should be found between both metallicity tracers
as long as the observed objects have a low nitrogen content $\left(\nicefrac{N}{H}<6.6\cdot10^{-6}\right)$
and show no  Wolf-Rayet features since the winds from
these stars could cause local nitrogen pollution. This conclusion
was confirmed by \cite{rosales-ortegaf.f.2006} using a sample of
71 low metallicity HII galaxies selected from the Sloan Digital Sky
Survey (SDSS) who also found, unexpectedly, that the $Y_{P}$ determination
using nitrogen was less  uncertain for his sample and showed a better agreement
with the WMAP experiment prediction than the corresponding value
found using oxygen. 

In this work we introduce the use of a third element, sulphur, as
metallicity tracer. Although less abundant than oxygen, sulphur is
also produced in massive stars through explosive nucleosynthesis and
its yield should closely follow that of O. Nebular S/H abundances
are therefore expected to follow those of O/H and the S/O ratio is
expected to remain constant at about the solar value, $logS/O\simeq-1.6$.
Empirical tests exploring this have been performed confirming this
ratio \citep{berg2015}.

Sulphur abundances can be measured from near infrared observations where the $\left[SIII\right]\lambda9069\AA$
and $\left[SIII\right]\lambda9532\AA$ lines appear among the most intense
features in the range. 
The region in the ionized nebulae where 
$S^{2+}$ originates  practically covers the $O^{+}$
and $O^{2+}$ one.
The corresponding electron temperature can be well
represented by $T_{e}\left[SIII\right]$ derived from the line intensity
ratio $I\left([SIII]\lambda\lambda9069\AA,9539\AA\right)/I\left([SIII]\lambda6312\right)$ \citep[e.g.][and references therein]{haegele2008}.
The use of the $S$ lines results in additional benefits compared
to the other elements. Firstly,  $\left[SIII\right]\lambda6312\AA$
falls in  a spectral region of high instrumental sensitivity. Secondly, these
lines are weakly affected by underlying stellar population absorption.
Furthermore, $\left[SII\right]$ lines can be measured with respect
to $H_{\alpha}$, while the near-infrared $\left[SIII\right]$ lines can be measured
with respect to nearby Paschen lines. This is useful to minimize reddening uncertainties 
and other calibration effects. Finally, and most important, the $S/H$
abundance can be measured by direct methods up to solar values. This
is crucial in order to increase the dynamical range  for the $\nicefrac{\Delta Y}{\Delta S}$
gradient calculation, with respect to the $O/H$ and $N/H$ metallicity
range. This would make the results less dependent on low $Y$ - metal poor
galaxies, while increasing the sample to objects with higher
$S/H$.

The paper has the following structure: Section 2 presents the observations
of the selected HII galaxy sample and the data
reduction. Section 3 describes the  analysis
of the data and the results attained. These results are discussed
in Section 4. Finally, the conclusions of this work are summarized
in Section 5.

\section{Observations and data reduction}

\label{Observations}

The HII galaxy sample was selected from the Sloan Digital Sky Survey (SDSS) to perform further ad-hoc observations.
The current Data release,
DR12, represents the final output from the SDSS-III survey (\cite{ahn2014}).
In addition to five photometric bands, spectroscopic data are included
for over $6.7\times10^{5}$ stars, galaxies and quasars. The light, covering 
 from $\ensuremath{3800\,\AA}$
to $9200\,\AA$ with a resolution power around $R=1800$ is received through a 
3-arcsec
diameter fiber aperture. For low
redshifts, this configuration enables an efficient selection of objects
with strong emission lines, including the $\left[SIII\right]\lambda9069\AA$
line.

The primordial helium abundance determination requires the analysis
of young, low metallicity and relatively high excitation nebulae,
which contain a negligible or small amount of neutral helium. Additionally,
a high signal-to-noise ratio is essential. High values of the hydrogen
emission line equivalent widths guarantee this requirement while also
selecting young objects with little underlying stellar population
absorption. The galaxy sample selection has then  been performed
following the criteria: 
\begin{itemize}
\item $Ew\left(H\alpha\right)>200\,\AA$ 
\item $0.7<\sigma_{H\alpha}<3.5\,\AA$
\item $\left(S/N\right)_{H\alpha}>100$ 
\item $\left[NII\right]\lambda6583/H\alpha<0.2$ 
\item $\left[OIII\right]\lambda4363/H\beta\geq0.1$ 
\item $0.0007<z<0.2$ 
\end{itemize}
Finally, BPT diagrams (\cite{baldwin1981}) were used in order to
exclude AGN and LINERs. Special care was taken not to include 
objects with wide Wolf-Rayet features: the blue bump, around the $HeII\lambda4686\AA$ line and the red bump, around the $CIV\lambda\,5808\AA$ line to avoid possible metal contamination. 

\subsection{Sample description}

\begin{figure*}
\noindent \begin{centering}
\includegraphics[width=0.99\textwidth]{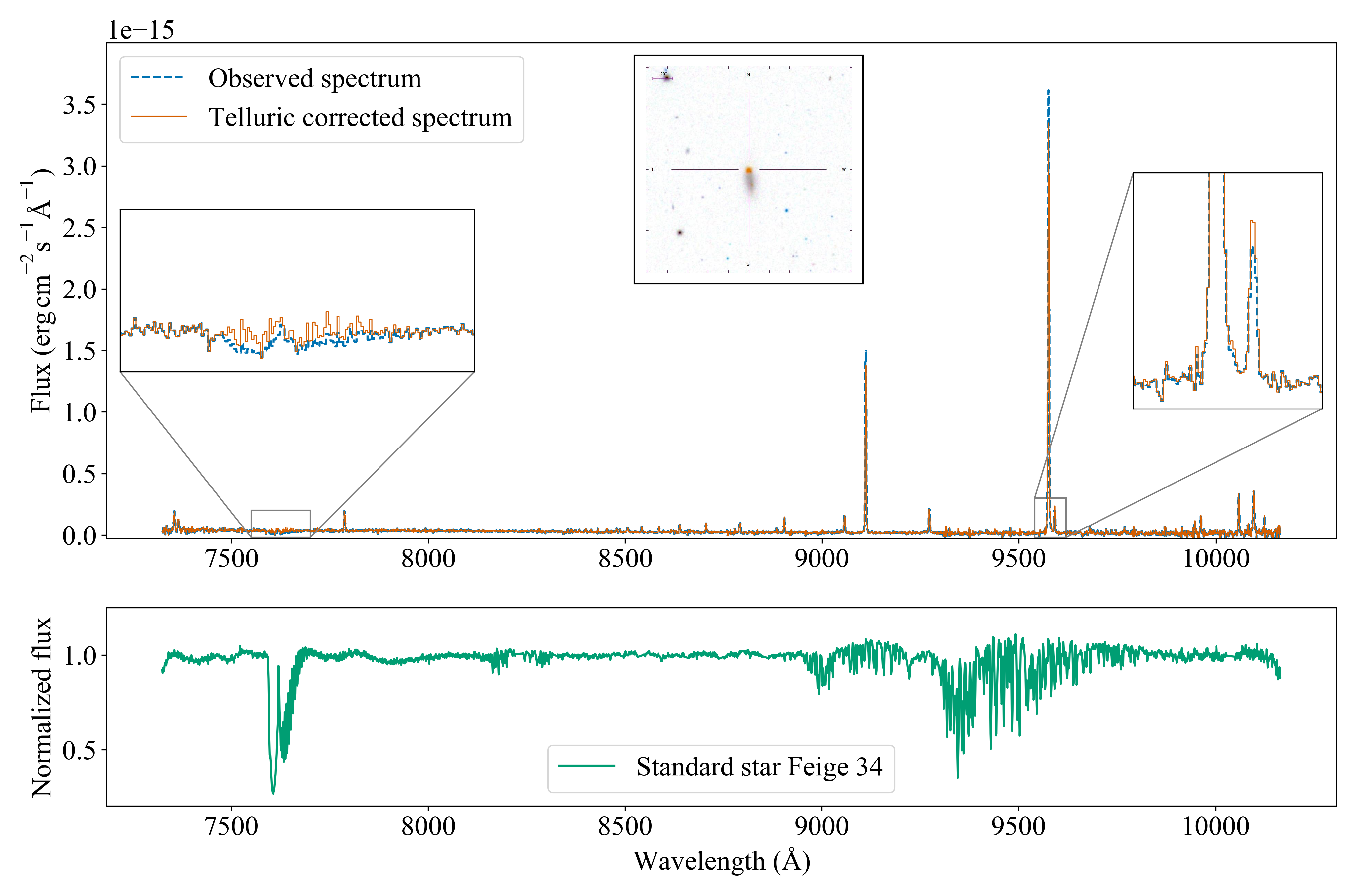}
\par\end{centering}
\caption{\label{fig:Telluric-correction-applied}Telluric correction applied
to SDSSJ024815.93-081716.5 red arm spectrum. The calibrated object
is divided by the normalized spectrum from a standard star whose absorption
features are artificially removed. The stellar spectrum was acquired
with the same slit width as the science objects.}
\end{figure*}
The  spectra were acquired with the double arm Intermediate Dispersion Spectrograph
and Imaging System (ISIS) attached to the 4.2m William Herschel Telescope
(WHT) at the Roque de los Muchachos Observatory during several observing
runs between 2008 and 2015. The observations were taken using two
different configurations.

\begin{table*}
\caption{\textcolor{black}{\label{tab:Observational-log}Sample observations.
Column (1) gives an existing object name (otherwise an order name
ad-hoc for this paper). Two regions (A1 and A2) were observed in MRK36.}}
\centering{}\begin{tabu}{lccccc}%
\hline%
Local reference&SDSS reference&z&RA (hh:':")&DEC (deg:':")&ISIS configuration\\%
\hline%
FTDTR-1&\href{http://dr12.sdss3.org/spectrumDetail?mjd=53725\&fiber=536\&plateid=2329}{SDSSJ012534.19+075924.5}&0.009593&1:25:34.212&+07:59:24.59&I\\%
IZw18&\href{http://dr12.sdss3.org/spectrumDetail?mjd=51991\&fiber=312\&plateid=556}{SDSSJ093402.02+551427.8}&0.002614&9:34:02.162&+55:14:22.34&I\\%
MRK36-A1&\href{http://dr12.sdss3.org/spectrumDetail?plateid=2211\&mjd=53786\&fiber=486}{SDSSJ110458.30+290816.6}&0.002190&11:04:58.564&+29:08:14.89&I\\%
MRK36-A2&\href{http://dr12.sdss3.org/spectrumDetail?plateid=2211\&mjd=53786\&fiber=486}{SDSSJ110458.30+290816.6}&0.002202&11:04:58.564&+29:08:14.89&I\\%
MRK475&\href{http://dr12.sdss3.org/spectrumDetail?plateid=1382\&mjd=53115\&fiber=175}{SDSSJ143905.46+364821.8}&0.002024&14:39:05.466&+36:48:21.97&I\\%
FTDTR-2&\href{http://dr12.sdss3.org/spectrumDetail?plateid=976\&mjd=52413\&fiber=600}{SDSSJ171236.63+321633.4}&0.01203&17:12:37.071&+32:16:29.52&I\\%
IZw70&\href{http://dr12.sdss3.org/spectrumDetail?plateid=1383\&mjd=53116\&fiber=110}{SDSSJ145056.56+353419.5}&0.003953&14:50:56.801&+35:34:16.74&I\\%
MRK689&\href{http://dr12.sdss3.org/spectrumDetail?plateid=1388\&mjd=53119\&fiber=17}{SDSSJ153619.43+304056.4}&0.005793&15:36:19.439&+30:40:56.40&I\\%
MRK67&\href{http://dr12.sdss3.org/spectrumDetail?plateid=2094\&mjd=53851\&fiber=487}{SDSSJ134156.47+303109.6}&0.003316&13:41:56.956&+30:31:05.91&I\\%
FTDTR-3&\href{http://dr12.sdss3.org/spectrumDetail?plateid=2233\&mjd=53845\&fiber=371}{SDSSJ122129.18+282208.3}&0.008695&12:21:29.178&+28:22:08.44&I\\%
SHOC022&\href{http://dr12.sdss3.org/spectrumDetail?plateid=418\&mjd=51817\&fiber=302}{SDSSJ003218.59+150014.1}&0.01793&0:32:19.337&+15:00:02.48&I\\%
FTDTR-4&\href{http://dr12.sdss3.org/spectrumDetail?plateid=861\&mjd=52318\&fiber=489}{SDSSJ081334.17+313252.1}&0.01959&8:13:33.374&+31:33:10.29&I\\%
SHOC220&\href{http://dr12.sdss3.org/spectrumDetail?plateid=550\&mjd=51959\&fiber=92}{SDSSJ084029.91+470710.2}&0.04210&8:40:28.462&+47:07:19.92&I\\%
FTDTR-5&\href{http://dr12.sdss3.org/spectrumDetail?plateid=942\&mjd=52703\&fiber=612}{SDSSJ100348.66+450457.6}&0.009252&10:03:47.356&+45:05:06.37&I\\%
FTDTR-6&\href{http://dr12.sdss3.org/spectrumDetail?plateid=456\&mjd=51910\&fiber=76}{SDSSJ024815.93-081716.5}&0.004528&2:48:16.920&-08:17:15.12&I\\%
FTDTR-7&\href{http://dr12.sdss3.org/spectrumDetail?plateid=1185\&mjd=52642\&fiber=123}{SDSSJ082334.84+031315.6}&0.009622&8:23:35.322&+03:13:17.57&I\\%
MRK627&\href{http://dr12.sdss3.org/spectrumDetail?plateid=934\&mjd=52672\&fiber=369}{SDSSJ084634.40+362620.8}&0.01059&8:46:33.891&+36:26:33.81&I\\%
SHOC592&\href{http://dr12.sdss3.org/spectrumDetail?plateid=640\&mjd=52200\&fiber=270}{SDSSJ212332.71-074831.0}&0.02805&21:23:32.225&-07:48:27.21&II\\%
PHL293B&\href{http://dr12.sdss3.org/spectrumDetail?plateid=376\&mjd=52143\&fiber=160}{SDSSJ223036.79-000636.9}&0.005138&22:30:36.429&-00:06:33.92&II\\%
SHOC588&\href{http://dr12.sdss3.org/spectrumDetail?plateid=639\&mjd=52146\&fiber=242}{SDSSJ211527.07-075951.3}&0.02837&21:15:27.200&-07:59:54.95&II\\%
SHOC036&\href{http://dr12.sdss3.org/spectrumDetail?plateid=394\&mjd=51913\&fiber=472}{SDSSJ005147.29+000940.0}&0.03738&0:51:47.394&+00:09:24.05&II\\%
SHOC575&\href{http://dr12.sdss3.org/spectrumDetail?plateid=358\&mjd=51818\&fiber=472}{SDSSJ172906.55+565319.3}&0.01601&17:29:07.491&+56:53:12.35&II\\%
SHOC579&\href{http://dr12.sdss3.org/spectrumDetail?plateid=358\&mjd=51818\&fiber=504}{SDSSJ173501.25+570308.5}&0.04739&17:35:00.687&+57:02:52.42&II\\%
FTDTR-8&\href{http://dr12.sdss3.org/spectrumDetail?plateid=942\&mjd=52703\&fiber=612}{SDSSJ100348.66+450457.6}&0.009251&10:03:48.766&+45:04:53.69&I\\%
SHOC263&\href{http://dr12.sdss3.org/spectrumDetail?plateid=556\&mjd=51991\&fiber=224}{SDSSJ093813.49+542825.1}&0.1022&9:38:13.761&+54:28:22.20&I\\%
FTDTR-9&\href{http://dr12.sdss3.org/spectrumDetail?plateid=501\&mjd=52235\&fiber=602}{SDSSJ100746.51+025228.3}&0.02330&10:07:46.500&+02:52:27.98&I\\%
FTDTR-10&\href{http://dr12.sdss3.org/spectrumDetail?plateid=575\&mjd=52319\&fiber=521}{SDSSJ102429.25+052450.9}&0.03313&10:24:29.237&+05:24:50.71&I\\%
\hline%
\end{tabu}
\end{table*}
In Configuration I, the R300B blue grating was centred at $\lambda_{c,\,Blue}=5950\AA$
covering between $4300\AA$ and $7600\AA$ and the red R316R one was
centred at $\lambda_{c,\,Red}=8655\AA$ to cover between $7260\AA$
and $10050\AA$. The 7500 dichroic was used in this set up.

Configuration II used the R300B grating covering between $3200\,\AA\,{\rm {and}\,5550\,\AA}$
and centred at $\lambda_{c,\,Blue}=4375\,\AA$. The R310R grating
in the red arm was centred at $\lambda_{c,\,Red}=7615\,\AA$ to cover 
between $5500\AA\,{\rm {and}\,9730\AA}$. The 5700 dichroic was used
for this configuration.

For both setups, the EEV12 and REDPLUS detectors
were used at the blue and red arm respectively. Both configurations
provided a similar spectral dispersion of $0.86\,\AA/pix$ with a
pixel size close to $0.2\,arcsec$. The slit width was adjusted to
the seeing with values between $1$ and $1.5\,arcsec$. The
observations were performed at small zenith distance and at parallactic
angle to avoid effects of differential refraction in the UV-blue.
In most cases these observations provided an adequate signal-to-noise
for the analysis of weak lines such as $\left[SII\right]\lambda\lambda6717\AA,6731\AA$,
$\left[OIII\right]\lambda4363\AA$, $\left[ArIV\right]\lambda4740\AA$ and
$\left[SIII\right]\lambda6312\AA$. Table \ref{tab:Observational-log}
presents the observed sample. The first and second columns give the
object name and the SDSS denomination respectively. The redshift is
shown in column 3. The object coordinates are given in columns 4 and
5 and the instrumental configuration used is shown in column 6.

Several bias and sky flat-field frames were obtained at both dusk
and dawn. Additionally, a pair of flatfields and one calibration lamp
exposure ($CuNe+CuAr$) were taken at each science position. The files
were processed with IRAF\footnote{IRAF: The Image Reduction and Analysis Facility is distributed by
the National Optical Astronomy Observatories, which is operated by
the Association of Universities for Research in Astronomy, Inc. (AURA)
under cooperative agreement with the National Science Foundation (NSF).} following standard procedures: bias subtraction, division by a normalized
flatfield and wavelength calibration. The latter was accomplished
using 20 to 40 lines. The spectra were flux calibrated and the edges
between blue and red arms matched.To this purpose, at least two spectrophotometric
standard stars were observed each night. This allowed a good flux
calibration with an estimated precision of around 5\%. 

\begin{figure*}
\includegraphics[width=0.99\textwidth]{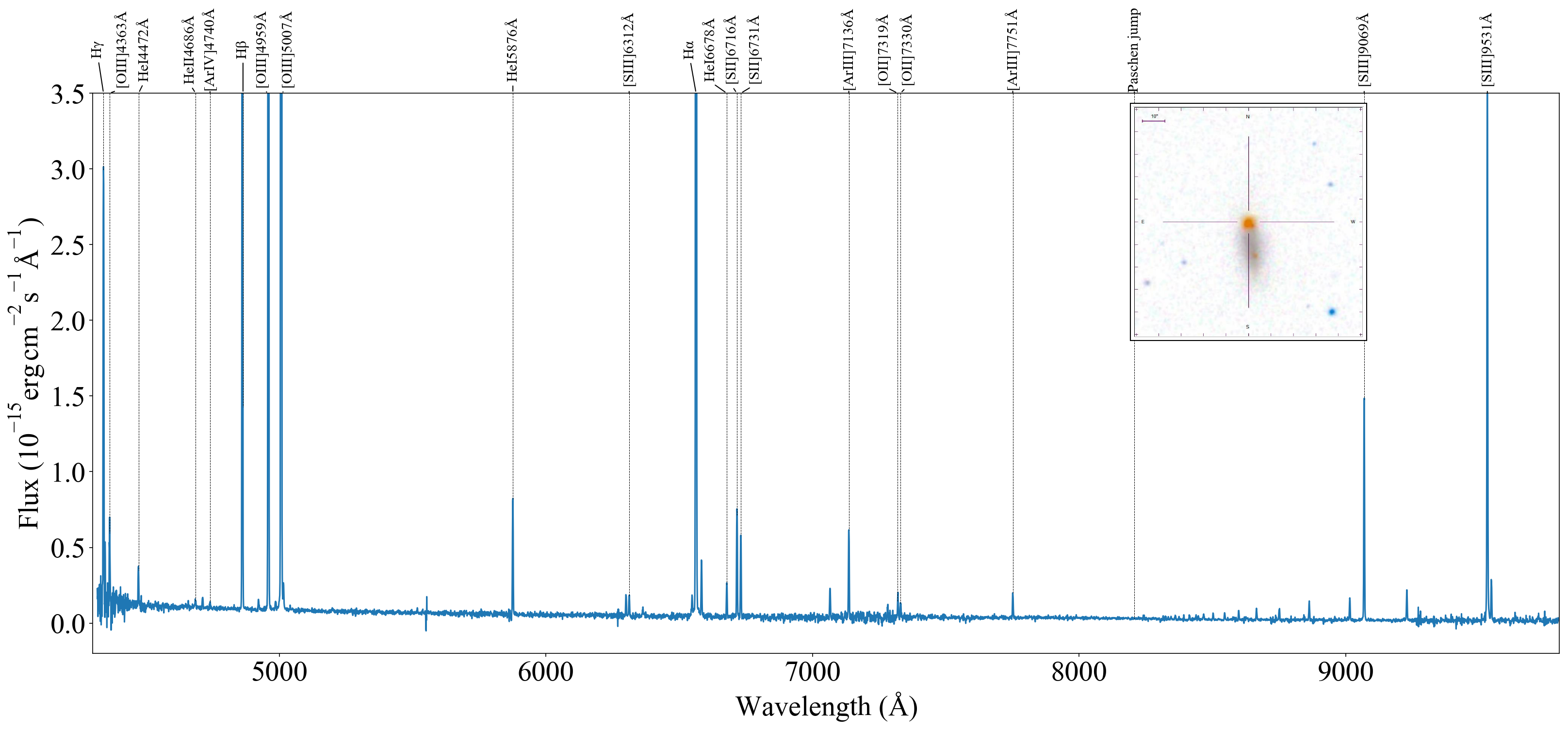}

\includegraphics[width=0.99\textwidth]{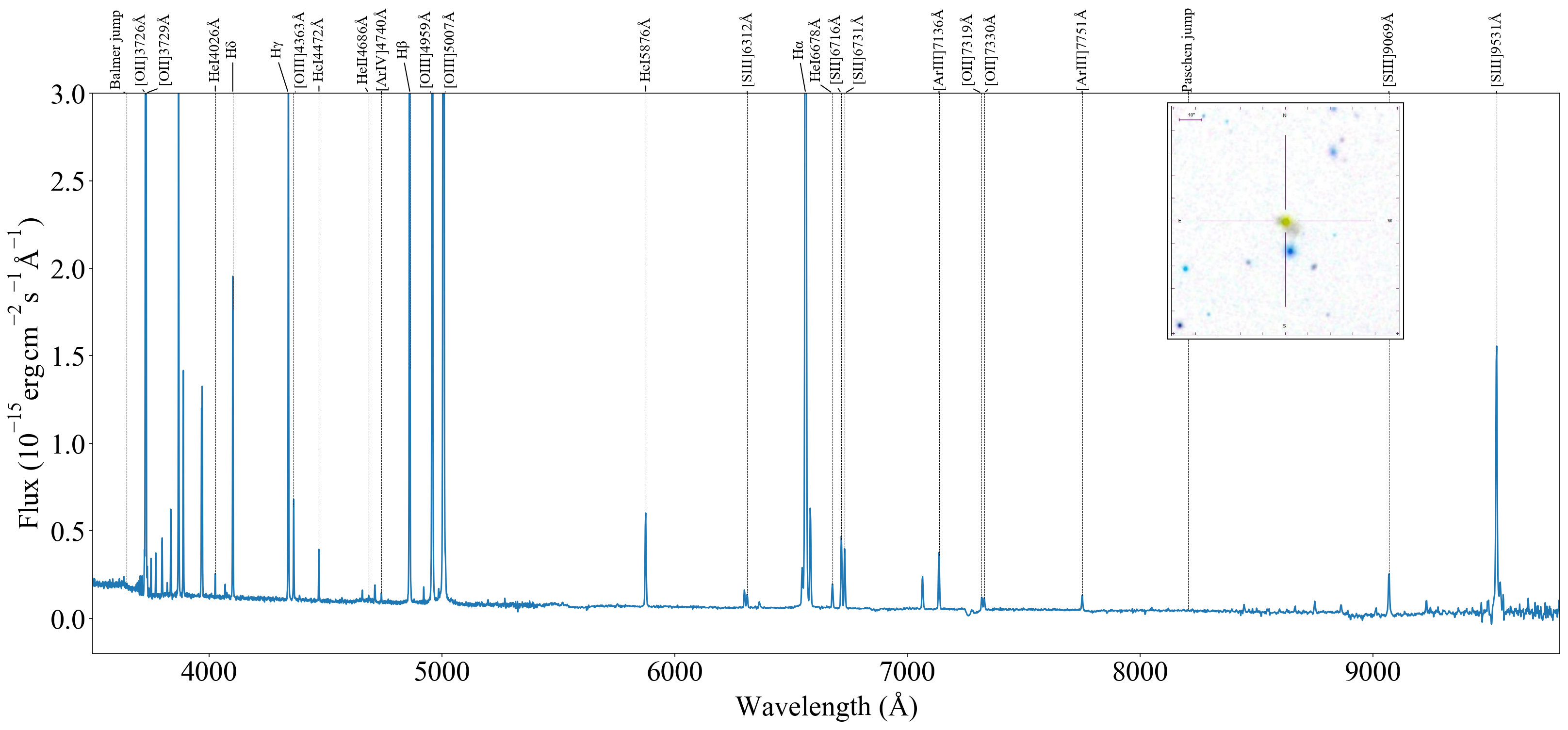}

\caption{\label{fig:Reduced_spectra}Sample of reduced spectra for objects
FTDR6 (top, configuration I) and SHOC579 (bottom, Configuration II).
Relevant lines are labelled. The object images belong
to the SDSS database.}
\end{figure*}

NIR observations are affected by atmospheric telluric contamination.
As an extra data reduction step, the object spectra were divided by
a normalized featureless spectrum from a standard star observed with
the same slit width as the object \citep{diaz1985}. Fig. \ref{fig:Telluric-correction-applied}
illustrates this procedure for SDSS J024815.93-081716.5. The normalized
spectrum belongs to the standard star Feige 34, whose absorption features
have been clipped prior to the normalization operation. Unfortunately,
the telluric observations could not be accomplished  every night.
For the objects whose $I(\left[SIII\right]\lambda9532\AA)/I(\left[SIII\right]\lambda9069\AA)$
ratio deviated considerably from the theoretical value \citep[2.47][]{hudson2012},
only one of the lines was considered in our analysis choosing the
one least affected by the atmospheric features. Two examples of the final
reduced spectra are shown in Fig.\ref{fig:Reduced_spectra}. The
remaining sample spectra can be seen in the manuscript supporting material.

\section{Analysis} 

\subsection{Spectral components}

The emission lines represent the most distinctive features in HII
galaxy spectra. The hydrogen and helium recombination lines are affected
by the absorption lines from the ionizing stars spectrum,  present
at the same wavelengths. These absorptions can be very important for
the fainter helium lines and must be taken into account in this element abundance
calculation. Moreover, in the youngest HII regions, the hydrogen
and helium nebular continua might contribute as much as the stellar
continuum. The following subsections describe how these three components:
nebular, stellar and emission line spectra are measured prior to undertaking
any chemical analysis.

\subsection{Hydrogen and helium nebular continua\label{subsec:Nebular-continuum-section}}

Three distinct components contribute to the nebular continuum; they
can be calculated from quantum physics principles. The free-free continuum
is governed by  Bremsstrahlung processes. As electrons travel within
the ionized hydrogen or helium Coulomb's radius, they suffer a deflection
and emit photons. These photons are of relatively low energy and hence
this component contributes mainly in the infrared spectral region.
This effect is calculated by \cite{brown1970} and \cite{osterbrock1974}:
{\footnotesize{}
\begin{eqnarray}
j_{\nu,\,} & = & j_{\nu,\,HI}+j_{\nu\,HeI}+j_{\nu\,HeII}\nonumber \\
 & = & \frac{1}{4\pi}n_{e}\left(n_{H^{+}}\gamma_{\nu}(H^{0})+n_{He^{+}}\gamma_{\nu}(He^{0})+\gamma_{\nu}(He^{+})\right)\label{eq:Emissivity}
\end{eqnarray}}
where $\left(j_{\nu}\right)$ is the continuum component emissivity,
$\left(n_{e}\right)$ is the electron density, $\left(n_{H},\,n_{He}\right)$
are the hydrogen and helium ion densities and $\left(\gamma_{\nu,\,ff}\right)$
is the continuous emission coefficient. In the case of the free-free
component, the latter coefficient can be calculated from the following
relation, in which the quantum effects can be neglected, $\left(g_{ff}\left(T_{e},Z,\nu\right)\approx1\right)$
for the electron temperatures commonly encountered in HII regions:
{\footnotesize{
\begin{eqnarray}
\gamma_{\nu,\,ff} & = & const.\left(\frac{\pi h\nu_{0}}{3kT_{e}}\right)^{\nicefrac{1}{2}}e^{-\nicefrac{h\nu}{kT}}
\end{eqnarray}
}
where $\nu_{0}$ denotes the threshold frequency for the ionization of
$H^{0}$ $\left(h\nu_{0}=13.6eV\right)$.

The free-bound emission, to which both hydrogen and helium atoms contribute,
is produced from the electron transitions from free to bound states
(recombination). This process is responsible for the Paschen and Balmer
jumps. Computing $\gamma_{\nu,\,fb}$, requires to add up all the
recombination to states of principal quantum number $(n)$ \citep[e.g.][]{ercolano2006}.
In their hydrogenic codes, convergence is obtained by considering
principal quantum levels up to $n=200$ for hydrogen and $n=350$
for helium. Their results are presented in tabular form, where $\gamma_{\nu,\,fb}$
is parametrised as a function of temperature and wavelength. 

Finally, in the ultraviolet region, the main nebular contribution
comes from the hydrogen $2s\rightarrow1s$ decay, which produces a
photon pair. Thereby, this emission is commonly referred to as 2-photon
continuum. The continuous emission coefficient $\left(\gamma_{\nu,\,2q}\right)$
is characterized by the parameter $g_{\nu}\left(T_{e}\right)$ \citep{nussbaumer1984}
following the scheme presented by \cite{osterbrock1974}:

\begin{equation}
\gamma_{\nu,2q}=\frac{\alpha_{eff}g_{\nu}}{1+\nicefrac{q_{2}}{A_{2q}}}
\end{equation}
where $\left(A_{2q}=8.23s\right)$ is the transition probability for
the photon decay and $\left(\alpha_{eff}\right)$ and $\left(q_{2}\right)$
are the effective recombination coefficient and collisional transition
rate for protons and electrons respectively. These parameters are
determined from the fits of \cite{molla2009}:

\begin{eqnarray}
\alpha_{eff} & = & 0.647\cdot10^{-10}\cdot T_{e}^{-0.722}\\
q_{2} & = & 5.92\cdot10^{-4}-6.1\cdot10^{-9}T_{e}
\end{eqnarray}

Once $\left(\gamma_{\nu}\right)$ has been tabulated for each process
and ion, the total nebular emissivity can be calculated from the $HeI$
and $HeII$ fractions. The final hydrogen and helium components are
computed as:

\begin{equation}
j_{\nu}=\frac{n_{e}n(H^{+})}{4\pi}\left(\gamma_{\nu}(H^{0})+\frac{n(He^{+})}{n(H^{0})}\gamma_{\nu}(He^{0})+\frac{n(He^{2+})}{n(H^{+})}\gamma_{\nu}(He^{+})\right)
\end{equation}
where, {\footnotesize{}
\begin{eqnarray}
\gamma_{\nu}(H^{0}) & = & \gamma_{\nu,ff}(H^{0})+\gamma_{\nu,fb}(H^{0})+\gamma_{\nu,2q}(H^{0})\\
\gamma_{\nu}(He^{0}) & = & \gamma_{\nu,ff}(He^{0})+\gamma_{\nu,fb}(H^{0})\\
\gamma_{\nu}(He^{+}) & = & \gamma_{\nu,ff}(He^{+})+\gamma_{\nu,fb}(H^{+})
\end{eqnarray}
}{\footnotesize \par}

Up to this point, the nebular continuum calculation requires some
estimates of the electron temperature and of $He^{0}$ and $He^{+}$
abundances. A characteristic low ionization temperature from the $\left[SIII\right]$
lines was used. This is a reasonable assumption as the twice ionized
sulphur population covers a large percentage of the HII region volume.

Translating emissivity to flux for a given object would require accurate
knowledge of both distance and density. Fortunately, a more feasible
approach is proposed by \cite{zanstrah.1931} to estimate the ultraviolet 
continuum produced by the ionizing stars from the recombination lines
intensity. In our case, we compare the emissivity from a recombination
line and the nebular continuum. From first principles, a line emission
luminosity, e.g.~$H\beta$, is given by: 
\begin{eqnarray}
L\left(H\beta\right) & = & h\nu_{H\beta}\int_{0}^{r_{1}}4\pi j_{H\beta}dV\label{eq:Hbeta_luminosityA}\\
 & = & h\nu_{H\beta}\int_{0}^{r_{1}}4\pi n_{p}n_{e}\alpha_{H\beta}^{eff}dV\label{eq:Hbeta_luminosityB}
\end{eqnarray}
where $\left(\alpha_{H\beta}^{eff}\right)$ is the effective recombination
coefficient for the $H\beta$ emission line which is tabulated in
\cite{pequignot1991}.

To obtain the luminosity from the nebular continuum we integrate the
emissivity $j_{\nu,}$ over the ionized volume:

\begin{eqnarray}
L_{\nu} & = & \int_{0}^{r_{1}}4\pi j_{\nu}dV=\int_{0}^{r_{1}}4\pi\left(\frac{1}{4\pi}n_{p}n_{e}\gamma_{\nu}\right)dV\\
 & = & \int_{0}^{r_{1}}n_{p}n_{e}\gamma_{\nu}dV\label{eq:Nebular_Luminosity}
\end{eqnarray}

Dividing eq. \ref{eq:Nebular_Luminosity} by eq.\ref{eq:Hbeta_luminosityA}
we obtain: 
\begin{equation}
\frac{L_{\nu}}{L\left(H\beta\right)}=\frac{\gamma_{\nu}\int_{0}^{r_{1}}n_{p}n_{e}dV}{h\nu_{H\beta}\alpha_{H\beta}^{eff}\int_{0}^{r_{1}}4\pi n_{p}n_{e}dV}
\end{equation}
which can be approximated by:

\begin{equation}
\frac{F_{Neb,\,\nu}}{F\left(H\beta\right)}\approx\frac{\gamma_{\nu}}{\alpha_{H\beta}^{eff}\cdot h\nu_{H\beta}}\label{eq:zanstra_calibration}
\end{equation}

This simplification is valid as long as the $\left(\gamma_{\nu}/\alpha_{H\beta}^{eff}\right)$
ratio remains constant with temperature. This is a reasonable assumption for the
physical conditions commonly encountered in HII galaxies.

\subsection{\label{subsec:Stellar-Continuum}Underlying stellar continuum}

The observed hydrogen and helium lines are the combination of the
emission lines from the ionized gas and the absorption lines from the ionizing stars
atmosphere. In order to obtain a pure emission line spectrum
the subtraction of the stellar contribution is mandatory. This can
be accomplished with the help of stellar population synthesis models.
Here we used the STARLIGHT synthesis code \citep{cidfernandes2005}.
In this code, we used a Single Stellar Population (SSP) library
with ages, between $1\,Myr$ and $17\,Gyr$, and metallicities,
between $Z_{\odot}/200$ and $1.5\cdot Z_{\odot}$. Suitable emission
line masks were manually selected for each observed spectrum as described
in the following section. Additionally, the first and final $50\,\AA$
of the observed spectra were also masked. The spectra were re-sampled
to $1\,\AA/pixel$ as required by the code.

In our analysis, the observations were reduced without taking into
account the error propagation. Consequently, the error per pixel is
not quantified. As an alternative, STARLIGHT computes a global value
from the flux rms used during the normalization. In our case,
this flux window is between $4750\AA$ and $4840\AA$. Initially
the minimum visual extinction was set to zero and the default maximum
velocity dispersion was set to each galaxy $\sigma_{CaII}$ or $\sigma_{\left[OIII\right]}$
value, the velocity dispersion measured from the width of the near-IR CaII triplet absorptions or from the $[OIII]$ emission lines.

\subsection{\label{subsec:Emission-Spectrum-1}Line emission}

Once the nebular and stellar continua are removed from the observed
spectrum, the pure emission line fluxes can be measured. A graphical
interface was designed to inspect the spectral features in the relevant
wavelength range. The detected emission line along with two adjacent
continuum regions of at least $20\,\AA$ were manually selected and
a linear continuum fit was performed. The integrated line flux was
then measured. 
The uncertainty from
each line flux and equivalent width was estimated using a bootstrap
algorithm: a set of random realizations of every line pixel was generated
using data poissonian $1\sigma$ uncertainty from the continua adjacent
to the line. At each iteration, the area under the line is measured
and a Gaussian fit performed. This generates a distribution for the
line flux and Gaussian parameters (peak intensity, central wavelength
and standard deviation) from which their $1\sigma$ uncertainty is
determined.

A similar procedure was applied to blended lines, where their flux
is determined assuming a combination of Gaussian curves. Extra care was
given to the $\left[NII\right]\lambda\lambda6548\AA,6583\AA$ and
$H\alpha$ region. Even though our resolution is high enough to separate
all three lines, for some objects $H\alpha$ has a broad component
that contaminates the $[NII]$ emission. The intensity and shape of
this component are commonly explained by gas inflows/outflows and
hence it changes from object to object. The following strategy provided the
best output: 
\begin{enumerate}
\item Three Gaussians were fitted for $\left[NII\right]6548\AA$, $H\alpha$,
$\left[NII\right]6583\AA$ narrow components with the following constraints:
the central wavelength of the narrow features must be within $\pm2\,\AA$
of the expected value; both hydrogen and nitrogen emissions are
assumed to share the same $\sigma$; and the $\left[NII\right]6583\AA/\left[NII\right]6548\AA$
flux ratio must be $2.94$ [assuming the $N^{+}$ atomic data from
\cite{tayal2011}, \cite{wiese1996} and \cite{galavis1997}]. 
\item The fitted narrow emission is removed from the spectrum leaving only
the $H\alpha$ broad component, which is fitted with a Gaussian curve. 
\item In the last step, both the narrow and wide $H\alpha$
components are fitted using constraints from the previous fits: the
wide component central wavelength is fixed from the value obtained
on the second fit. Additionally, its peak flux is also contained within 
$\pm10\%$ the value obtained in the previous fitting. The narrow
components have the same constraints as those described for the first
fitting.
\end{enumerate}

\subsection{Chemical analysis\label{sec:Chemical-analysis}}

\begin{table*}
\caption{\label{tab:lines fluxes}Relative line intensities with respect to $H\beta$ and multiplied by 1000 for objects FTDTR-6
(Configuration I) and SHOC579 (Configuration II). Nebular and stellar
continua have been removed.}

{\scriptsize{}\begin{tabu}{lcccccc}%
&&FTDTR-6&&&SHOC579&\\%
$\lambda(\AA)$&$EW(\AA)$&$F(\lambda)$&$I(\lambda)$&$EW(\AA)$&$F(\lambda)$&$I(\lambda)$\\%
\hline%
3688 $H19$&-&-&-&0.4$\pm$0.07&3.080$\pm$0.50&3.143$\pm$0.51\\%
3705 $H16$&-&-&-&2$\pm$0.3&14.50$\pm$1.1&14.80$\pm$1.1\\%
3713 $H15$&-&-&-&2$\pm$0.3&10.73$\pm$1.3&10.94$\pm$1.4\\%
3726 $[OII]$&-&-&-&56.0$\pm$4.88&347.6$\pm$8.2&354.5$\pm$8.3\\%
3729 $[OII]$&-&-&-&65.8$\pm$5.65&409.1$\pm$7.1&417.2$\pm$7.3\\%
3750 $H12$&-&-&-&5$\pm$0.4&28.32$\pm$1.1&28.87$\pm$1.2\\%
3770 $H11$&-&-&-&5.67$\pm$0.413&33.02$\pm$0.81&33.64$\pm$0.82\\%
3797 $H10$&-&-&-&7.91$\pm$0.531&47.02$\pm$0.90&47.89$\pm$0.92\\%
3836 $H9$&-&-&-&10.8$\pm$0.600&64.64$\pm$0.74&65.78$\pm$0.76\\%
3869 $[NeIII]$&-&-&-&67.8$\pm$3.95&407.7$\pm$1.0&414.6$\pm$1.0\\%
3890 $H8$&-&-&-&28.5$\pm$1.86&170.8$\pm$0.97&173.6$\pm$0.99\\%
3968 $[NeIII]$&-&-&-&22.7$\pm$1.36&129.6$\pm$4.8&131.5$\pm$4.8\\%
3970 $H\epsilon$&-&-&-&25.7$\pm$1.39&146.9$\pm$3.9&149.1$\pm$4.0\\%
4026 $HeI$&-&-&-&3$\pm$0.2&16.24$\pm$0.78&16.46$\pm$0.80\\%
4069 $[SII]$&-&-&-&2$\pm$0.2&10.09$\pm$0.81&10.22$\pm$0.82\\%
4076 $[SII]$&-&-&-&0.8$\pm$0.1&4.123$\pm$0.69&4.177$\pm$0.70\\%
4102 $H\delta$&-&-&-&44.9$\pm$2.68&240.4$\pm$0.98&243.5$\pm$1.0\\%
4340 $H\gamma$&84.8$\pm$54.7&476.1$\pm$9.1&491.2$\pm$9.5&90.3$\pm$4.42&441.9$\pm$0.74&445.7$\pm$0.76\\%
4363 $[OIII]$&14.3$\pm$6.75&77.45$\pm$5.6&79.81$\pm$5.8&15.9$\pm$0.719&76.59$\pm$0.78&77.23$\pm$0.78\\%
4388 $HeI$&-&-&-&1$\pm$0.1&4.767$\pm$0.57&4.805$\pm$0.57\\%
4471 $HeI$&7.43$\pm$1.23&38.37$\pm$2.1&39.27$\pm$2.1&8.67$\pm$0.628&39.30$\pm$1.0&39.56$\pm$1.0\\%
4659 $[FeIII]$&0.8$\pm$0.2&3.904$\pm$0.78&3.949$\pm$0.79&2$\pm$0.2&9.156$\pm$0.75&9.186$\pm$0.75\\%
4686 $HeII$&2$\pm$0.3&10.35$\pm$1.0&10.45$\pm$1.1&1$\pm$0.2&4.753$\pm$0.87&4.766$\pm$0.87\\%
4711 $[ArIV]$&3$\pm$0.2&11.87$\pm$0.63&11.97$\pm$0.63&3$\pm$0.3&14.90$\pm$0.62&14.93$\pm$0.62\\%
4740 $[ArIV]$&1$\pm$0.2&5.240$\pm$0.74&5.276$\pm$0.75&2$\pm$0.2&6.456$\pm$0.66&6.469$\pm$0.66\\%
4861 $H\beta$&243$\pm$18.0&1000$\pm$0.00&1000$\pm$0.0&240$\pm$12.1&1000$\pm$0.0&1000$\pm$0.0\\%
4922 $HeI$&3$\pm$0.3&11.32$\pm$0.81&11.28$\pm$0.80&3$\pm$0.2&11.81$\pm$0.59&11.80$\pm$0.59\\%
4959 $[OIII]$&479$\pm$42.1&1970$\pm$2.0&1960$\pm$2.2&423$\pm$55.5&2039$\pm$2.4&2036$\pm$2.4\\%
4987 $[FeIII]$&2$\pm$0.2&9.587$\pm$0.66&9.524$\pm$0.65&2$\pm$0.2&9.306$\pm$0.65&9.289$\pm$0.65\\%
5007 $[OIII]$&1460$\pm$138&6003$\pm$140&5957$\pm$140&1290$\pm$68.1&6233$\pm$3.8&6220$\pm$3.8\\%
5016 $HeI$&7.05$\pm$27.1&29.00$\pm$110&28.77$\pm$110&-&-&-\\%
5198 $[NI]$&-&-&-&1$\pm$0.3&4.714$\pm$0.82&4.692$\pm$0.81\\%
5518 $[ClIII]$&1$\pm$0.3&3.942$\pm$0.92&3.819$\pm$0.89&1$\pm$0.2&4.903$\pm$0.59&4.860$\pm$0.58\\%
5755 $[NII]$&-&-&-&0.9$\pm$0.2&2.856$\pm$0.49&2.823$\pm$0.48\\%
5876 $HeI$&46.7$\pm$7.44&119.8$\pm$1.0&114.2$\pm$1.1&44.8$\pm$1.57&138.0$\pm$0.53&136.2$\pm$0.52\\%
6300 $[OI]$&10.2$\pm$2.72&21.72$\pm$1.3&20.36$\pm$1.2&9.46$\pm$0.438&26.27$\pm$0.58&25.80$\pm$0.57\\%
6312 $[SIII]$&10.8$\pm$2.92&22.21$\pm$1.6&20.81$\pm$1.5&6.53$\pm$0.306&18.18$\pm$0.46&17.85$\pm$0.45\\%
6364 $[OI]$&5$\pm$0.9&9.344$\pm$0.83&8.739$\pm$0.78&4$\pm$0.2&9.867$\pm$0.41&9.685$\pm$0.40\\%
6548 $[NII]$&9.35$\pm$2.17&19.21$\pm$0.017&17.85$\pm$0.13&20.3$\pm$0.859&58.93$\pm$0.55&57.74$\pm$0.54\\%
6563 $H\alpha$&1580$\pm$367&3244$\pm$33&3012$\pm$38&916$\pm$53.1&2658$\pm$1.8&2604$\pm$2.7\\%
6583 $[NII]$&27.5$\pm$14.7&56.48$\pm$27&52.41$\pm$25&50.6$\pm$2.41&150.3$\pm$0.74&147.2$\pm$0.73\\%
6678 $HeI$&17.0$\pm$4.41&32.76$\pm$1.1&30.31$\pm$1.0&13.6$\pm$0.756&36.64$\pm$0.61&35.86$\pm$0.60\\%
6716 $[SII]$&60.5$\pm$15.3&116.6$\pm$1.4&107.8$\pm$1.5&41.7$\pm$1.24&110.0$\pm$0.42&107.6$\pm$0.42\\%
6731 $[SII]$&45.0$\pm$10.8&86.13$\pm$1.4&79.55$\pm$1.4&33.2$\pm$0.923&87.19$\pm$0.39&85.29$\pm$0.39\\%
7065 $HeI$&20.0$\pm$7.72&32.04$\pm$1.5&29.29$\pm$1.4&20.6$\pm$0.835&50.87$\pm$0.50&49.62$\pm$0.49\\%
7136 $[ArIII]$&55.2$\pm$25.4&95.15$\pm$2.2&86.80$\pm$2.2&35.7$\pm$1.51&87.04$\pm$0.57&84.86$\pm$0.56\\%
7281 $HeI$&6.16$\pm$3.92&9.439$\pm$2.3&8.576$\pm$2.1&-&-&-\\%
7319 $[OII]$&14.5$\pm$6.49&23.66$\pm$2.0&21.47$\pm$1.8&13.2$\pm$1.33&24.54$\pm$1.0&23.89$\pm$1.0\\%
7330 $[OII]$&10.1$\pm$4.56&16.50$\pm$1.8&14.97$\pm$1.6&12.0$\pm$1.24&22.33$\pm$1.1&21.74$\pm$1.0\\%
7751 $[ArIII]$&15.8$\pm$2.86&24.98$\pm$0.79&22.42$\pm$0.75&12.6$\pm$0.809&24.77$\pm$0.61&24.04$\pm$0.59\\%
8392 $Hpa20$&3$\pm$0.5&2.634$\pm$0.31&2.330$\pm$0.27&-&-&-\\%
8413 $Hpa19$&5$\pm$1&4.694$\pm$0.47&4.151$\pm$0.42&0.9$\pm$0.4&1.587$\pm$0.66&1.534$\pm$0.63\\%
8438 $Hpa18$&4$\pm$0.8&4.216$\pm$0.38&3.726$\pm$0.34&-&-&-\\%
8446 $HeI$&4$\pm$0.7&4.288$\pm$0.37&3.789$\pm$0.33&5.74$\pm$0.643&9.535$\pm$0.66&9.214$\pm$0.64\\%
8467 $Hpa17$&6.35$\pm$1.13&6.695$\pm$0.53&5.913$\pm$0.48&-&-&-\\%
8502 $Hpa16$&5.40$\pm$0.921&5.664$\pm$0.38&4.999$\pm$0.34&-&-&-\\%
8545 $Hpa15$&6.16$\pm$1.13&6.389$\pm$0.47&5.634$\pm$0.42&-&-&-\\%
8598 $Hpa14$&8.62$\pm$2.70&8.383$\pm$0.71&7.385$\pm$0.63&2$\pm$0.8&3.807$\pm$1.1&3.675$\pm$1.0\\%
8665 $Hpa13$&11.4$\pm$2.23&11.31$\pm$0.49&9.950$\pm$0.45&5.23$\pm$1.15&8.375$\pm$1.2&8.082$\pm$1.2\\%
8750 $Hpa12$&14.6$\pm$3.42&12.87$\pm$0.54&11.30$\pm$0.50&11.0$\pm$1.32&17.08$\pm$0.82&16.48$\pm$0.79\\%
8863 $Hpa11$&17.8$\pm$3.47&17.38$\pm$0.50&15.23$\pm$0.48&11.3$\pm$2.07&15.06$\pm$1.1&14.52$\pm$1.1\\%
9015 $Hpa10$&24.9$\pm$6.66&22.12$\pm$0.61&19.33$\pm$0.59&13.1$\pm$3.19&11.55$\pm$1.0&11.13$\pm$0.97\\%
9069 $[SIII]$&251$\pm$62.0&225.8$\pm$0.62&197.2$\pm$2.7&106$\pm$35.1&84.91$\pm$1.9&81.78$\pm$1.8\\%
9229 $Hpa9$&37.9$\pm$11.5&31.87$\pm$0.70&27.75$\pm$0.71&15.8$\pm$5.88&20.62$\pm$2.4&19.85$\pm$2.3\\%
9531 $[SIII]$&815$\pm$465&591.5$\pm$1.2&512.4$\pm$7.2&250$\pm$124&493.6$\pm$5.7&474.4$\pm$5.5\\%
9546 $Hpa8$&55.3$\pm$33.4&44.40$\pm$1.5&38.46$\pm$1.4&22.8$\pm$8.89&51.75$\pm$3.9&49.73$\pm$3.8\\%
10012 $[OII]$&97.1$\pm$129&55.03$\pm$2.3&47.33$\pm$2.1&-&-&-\\%
\hline%
$I(H\beta)$&&2.30e-14$\pm$2.00e-17&2.90e-14$\pm$6.40e-16&&2.20e-14$\pm$1.30e-17&2.40e-14$\pm$6.10e-17\\%
$(erg\,cm^{-2} s^{-1} \AA^{-1})$&&&&&&\\%
$c(H\beta)$&&0.10$\pm$0.01&&&0.028$\pm$0.001&\\%
\hline%
\hline%
\end{tabu}}{\scriptsize \par}
\end{table*}
Table \ref{tab:lines fluxes} includes the emission line data for
two example HII galaxies, one for each configuration. Column (1) lists
the line and wavelength. Column (2) provides the equivalent width.
Column (3) and (4) give the line flux and the reddening-corrected
line intensity in units of $H\beta=1000$. Finally, at the foot of
the table, the raw and reddening corrected $H\beta$ fluxes for both galaxies are given in $erg\,cm^{-2}\,s^{-1}\,\AA^{-1}$.
These values have been corrected for the underlying stellar continuum.
The remaining tables for the complete sample are available online. The emission lines were corrected for reddening. The $c\left(H\beta\right)$
coefficient, determined canonically from the Balmer decrement is also
given in the table. We considered appropriate for these young starforming
bursts the ``LMC average'' reddening curve from \cite{gordon2003}
with $R_{V}=3.4$. The recombination coefficients
are calculated using PyNeb \citep[see][among others]{luridiana2015} with the atomic data
of \cite{storey1995}. The Paschen series was not included in the
$c\left(H\beta\right)$ calculation as the simple stellar populations
used reach only up to $7000\AA$ as already mentioned.

The abundances of He, S, O and N are calculated
from the reddening corrected emission line fluxes also using PyNeb.
The tools provided by this library were scripted into a Monte Carlo
algorithm to propagate the uncertainty in the emission line fluxes
to the electron temperatures and densities, and finally, to the chemical
abundances. The abundances were determined following these
steps: 
\begin{enumerate}
\item PyNeb calculates simultaneously $n_{e}$ and $T_{e}$ using suitable
density and temperature diagnostic lines. Lines which arise from similar
excitation energies for the former and the ratio of lines with different
excitation energies to estimate the latter as usual. The $S^{+}$
density determined from the $\left[SII\right]\mbox{\ensuremath{\lambda\lambda}}6716\AA$,
$6731\AA$ lines diagnostic is adopted for all species. This diagnostic,
however, starts to loose sensitivity at densities below $100\,cm^{-3}$.
For objects where the density calculated from these diagnostic was below
$75\,cm^{-3}$ the value was set to $50\pm25\,cm^{-3}$. 
\item We considered a three zone model for the electron temperature:
a high ionization zone where ions like $O^{2+}$, $Ne^{2+}$ are produced;
a lower ionization zone where ions like $O^{+}$, $Ne^{+}$, $N^{+}$,
$S^{+}$ are formed and a third intermediate ionization zone which
partially overlaps the previous two ones, where ions like $S^{2+}$
and $Ar^{2+}$ are formed. The different temperatures that characterize
each ion population are further discussed in the following subsections. 
\item For the majority of the objects, similar temperature
precision was attained for $T_{e}\left[SIII\right]$ and $T_{e}\left[OIII\right]$.
This can be noticed in Fig. \ref{fig:Temperature-comparison-O-S},
where both temperatures are plotted for the observed galaxies.
For some objects though, the uncertainty is larger for one of the
temperatures. If this is the $T_{e}\left[OIII\right]$, it usually
corresponds to data obtained with  instrument configuration I, in which the $\left[OIII\right]\lambda4363\AA$
line is very close to the blue edge of the spectrum,  hence it is 
noisier. The cases with larger error in $T_{e}\left[SIII\right]$ are
those that show important telluric contamination around the $S^{2+}$
lines, where the continuum was extracted. Still, as seen in the figure,
the correspondence is quite good as expected \citep{garnett1992}.
Consequently, it is possible to replace the higher-uncertainty temperature
using the linear fitting, while propagating the lower uncertainty.
We used eq. \ref{eq:TSIII_TOIII_relation}, taken from \cite{perez-montero2005},
who use the same atomic parameters as we are using in this work for the sulphur and oxygen ions: 
\begin{equation}
t_{e}\left[OIII\right]=1.0807\times t_{e}\left[SIII\right]-0.0846\label{eq:TSIII_TOIII_relation}
\end{equation}
Additional literature fits are also plotted in Fig. \ref{fig:Temperature-comparison-O-S},
as it can be appreciated, they are all very similar in the $10000$ -
$16000\,K$ range.
\item The ionic abundances are calculated from the most intense line (if
more than one is available their intensities are added up). 
\item For the error propagation we used a bootstrap algorithm: for each
line a one thousand length array is generated. The fluxes in this
array come from a normal distribution defined by the mean line intensity
and its uncertainty. 
\item Since the bootstrap algorithm assumes a normal distribution for the
generated flux arrays, some diagnostic ratios may lie beyond the physical
boundaries. In these cases, PyNeb generates a non numerical entry.
These values are systematically removed at each step and are replaced
with random values generated from a normal distribution of the valid
entries. 
\end{enumerate}

\begin{figure}
\begin{centering}
\includegraphics[width=0.5\textwidth]{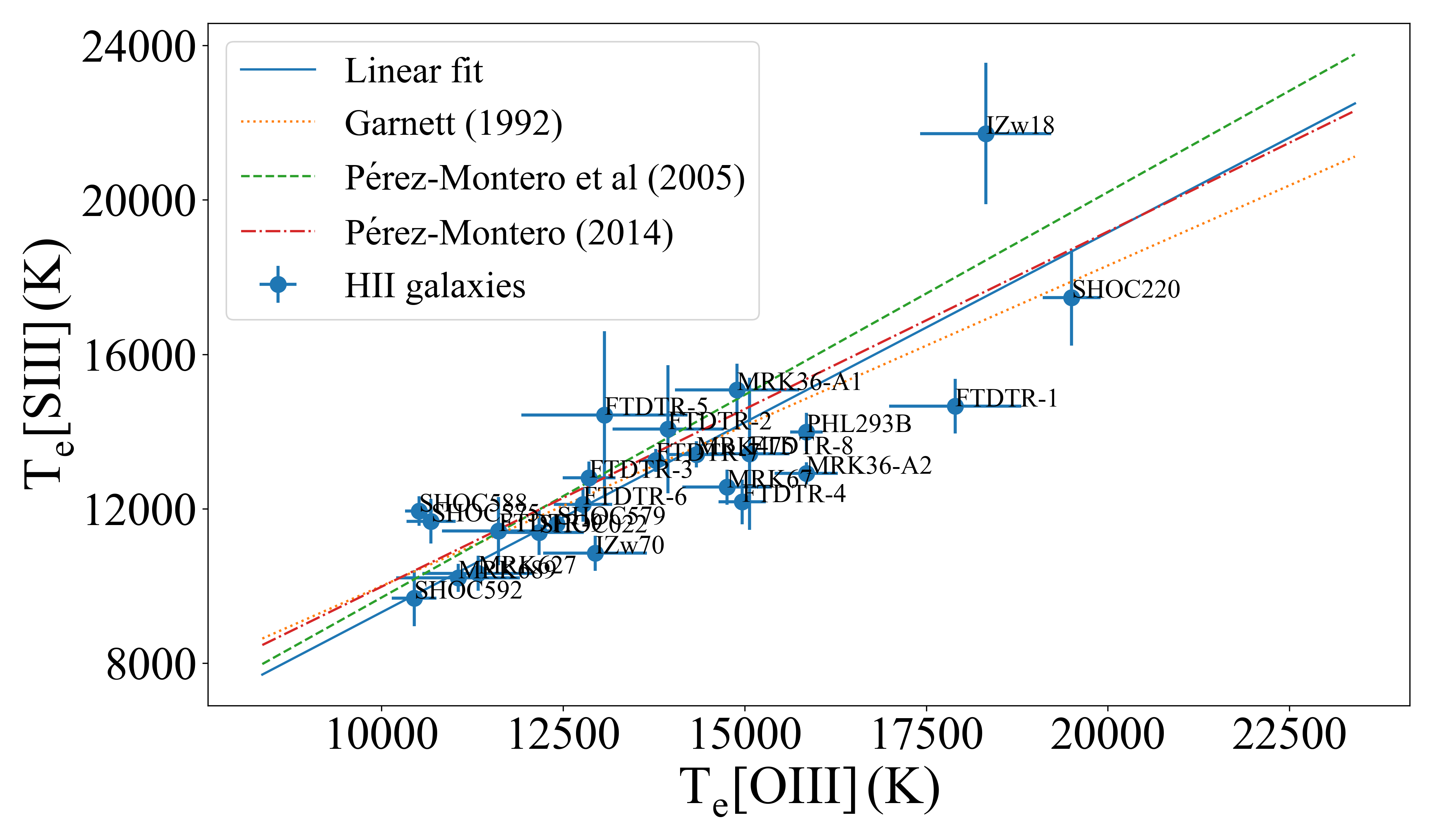}
\par\end{centering}
\caption{\label{fig:Temperature-comparison-O-S}T$_{e}\left[SIII\right]$ vs.
T$_{e}\left[OIII\right]$. The blue solid line is the fit to the data.
The dotted orange line represents the models prediction by Garnett
(1992). The green and red dashed lines correspond to the photoionization
models from P\'erez-Montero et al. (2005) and P\'erez-Montero (2014) respectively. }
\end{figure}

\begin{table*}
\caption{\label{tab:Atomic-data-references}Atomic data sources}
\centering{}%
\begin{tabular}{ccc}
\hline 
Ion & \multicolumn{2}{c}{Atomic data}\tabularnewline
\hline 
\hline 
$H$ & \multicolumn{2}{c}{\cite{storey1995}}\tabularnewline
\hline 
$He$ & \multicolumn{2}{c}{\cite{porter2013}}\tabularnewline
\hline 
$He^{+}$ & \multicolumn{2}{c}{\cite{storey1995}}\tabularnewline
\hline 
\hline 
Ion & Collision Strengths & Transition probabilities\tabularnewline
\hline 
\hline 
$O^{+}$ & \cite{pradhan2006,tayal2007} & \cite{zeippen1982,wiese1996}\tabularnewline
\hline 
$O^{+2}$ & \cite{aggarwal2000} & \cite{storey2000,wiese1996}\tabularnewline
\hline 
$N^{+}$ & \cite{tayal2011} & \cite{wiese1996,galavis1997}\tabularnewline
\hline 
$S^{+}$ & \cite{tayal2010} & \cite{podobedova2009}\tabularnewline
\hline 
$S^{+2}$ & \textcolor{black}{\cite{hudson2012}} & \cite{podobedova2009}\tabularnewline
\hline 
$S^{+3}$ & \cite{tayal2000} & \cite{dufton1982,johnson1986}\tabularnewline
\hline 
$Ar^{+2}$ & \cite{galavis1995} & \cite{kaufman1986,galavis1995}\tabularnewline
\hline 
$Ar^{+3}$ & \cite{ramsbottom1997} & \cite{mendoza1982}\tabularnewline
\hline 
\end{tabular}
\end{table*}
The references for the atomic data used in these calculations are
listed in Table \ref{tab:Atomic-data-references}.

\subsubsection{\label{subsec:Helium-abundance}Helium:}

The helium spectrum shows several lines within the observed range. Clearly,
the greater the number of lines considered, the most robust the analysis.
However, not all the lines share the same dependence on $T_{e}$
and $n_{e}$ and some are faint. Moreover, not all lines are evenly
affected by external physical phenomena e.g.~bluer lines are more
susceptible to underlying stellar absorption and reddening.
\citep[e.g.][and references therein]{aver2015} solved simultaneously for
the six physical parameters contributing to the helium line intensity.
In the present study these are quantified separately. The stellar absorption
is corrected from the SSPs fit once the nebular continuum has been
removed from the observed spectrum. The reddening coefficient $\left(c\left(H\beta\right)\right)$
is calculated from the Balmer decrement. The collisional effect on 
the helium lines is already included in the emissivities 
 \citep{porter2013}. Finally, we assumed $T\left(HeI\right)=T_{e}\left[OIII\right]=T_{high}$
and $n_{e}\left(HeI\right)=n_{e}\left[SII\right]$. At this point
the abundance is calculated using $\chi^{2}$ fitting over the set of $HeI$
lines: 
\begin{equation}
\chi^{2}=\sum_{i=1}^{n}\frac{\left(I_{He\lambda,\,i,\,obs}-I_{He\lambda,\,i,theo}\right)^{2}}{\sigma^{2}_{He\lambda,\,i}}
\end{equation}
where $\left(I_{He\lambda,\,i}\right)$ corresponds to a particular
$HeI$ line intensity and $\left(\sigma_{i}\right)$ to its uncertainty
(both normalized to $H\text{\ensuremath{\beta}}=1000$). The theoretical
line intensity is computed as: 
\begin{equation}
I_{He\lambda,\,i,theo}=y^{+}\cdot\frac{E_{He\lambda}\left(T_{high},\,n_{e}[SII]\right)}{E_{\beta}\left(T_{High},\,n_{e}[SII]\right)}
\end{equation}

This fit is repeated in a one thousand bootstrap algorithm, in which $T_{e}$
and $n_{e}$ values are randomly generated from the measured values assuming a Gaussian distribution. The algorithm output is itself a distribution
which provides $He^{+}$ mean value and uncertainty. The code uses
the HeI $\lambda4471\AA,5876\AA,6678\AA$ lines to perform the fit.

A similar approach is used to calculate the $He^{2+}$
abundance if the $HeII\lambda4686\mathring{A}$ line is observed.
The previous assumptions imply that the HeI and HeII emission are
produced in the high ionization region. This is an acceptable assumption
for $HeII$ as can be seen by using photoionization models. For HeI, Peimbert and colaborators \citep{peimbert2002}
analysed the possible bias introduced on the primordial helium determination
by assuming $T_{e}\left(HeI\right)=T_{e}\left(OIII\right)$. However,
as discussed by \citet{aver2010} and \citet{aver2015}, a
large degeneracy in the temperature exists when calculating helium
abundance, quite possibly enhanced by temperature fluctuations \citep{luridiana2003}.
To overcome this degeneracy, Aver and collaborators have used a temperature
prior based on the measured T$_{e}([OIII])$ to fit the model parameters.
Still, the mean uncertainty in the helium temperature for their flagged
sample is $\Delta T_{e}\left(H_{e}\right)\approx2200\,K$. The approach
proposed here represents a simplification that consists on using
$T_{e}([OIII])$ to generate a distribution of values to compute the
$T_{e}(HeI).$ A more rigorous estimate of $T_{e}(HeI)$ is left for
a forthcoming paper (Fern\'andez et al, in preparation). The helium
lines observed in both our instrumental setups are shown in Table
\ref{tab:lines fluxes}. The total He abundance is calculated as:
\begin{equation}
\frac{He}{H}=\frac{He^{+}}{H^{+}}+\frac{He^{2+}}{H^{+}}
\end{equation}

\subsubsection{\label{subsec:sulphur abundance}Sulphur}

\begin{figure*}
\begin{centering}
\includegraphics[width=\textwidth]{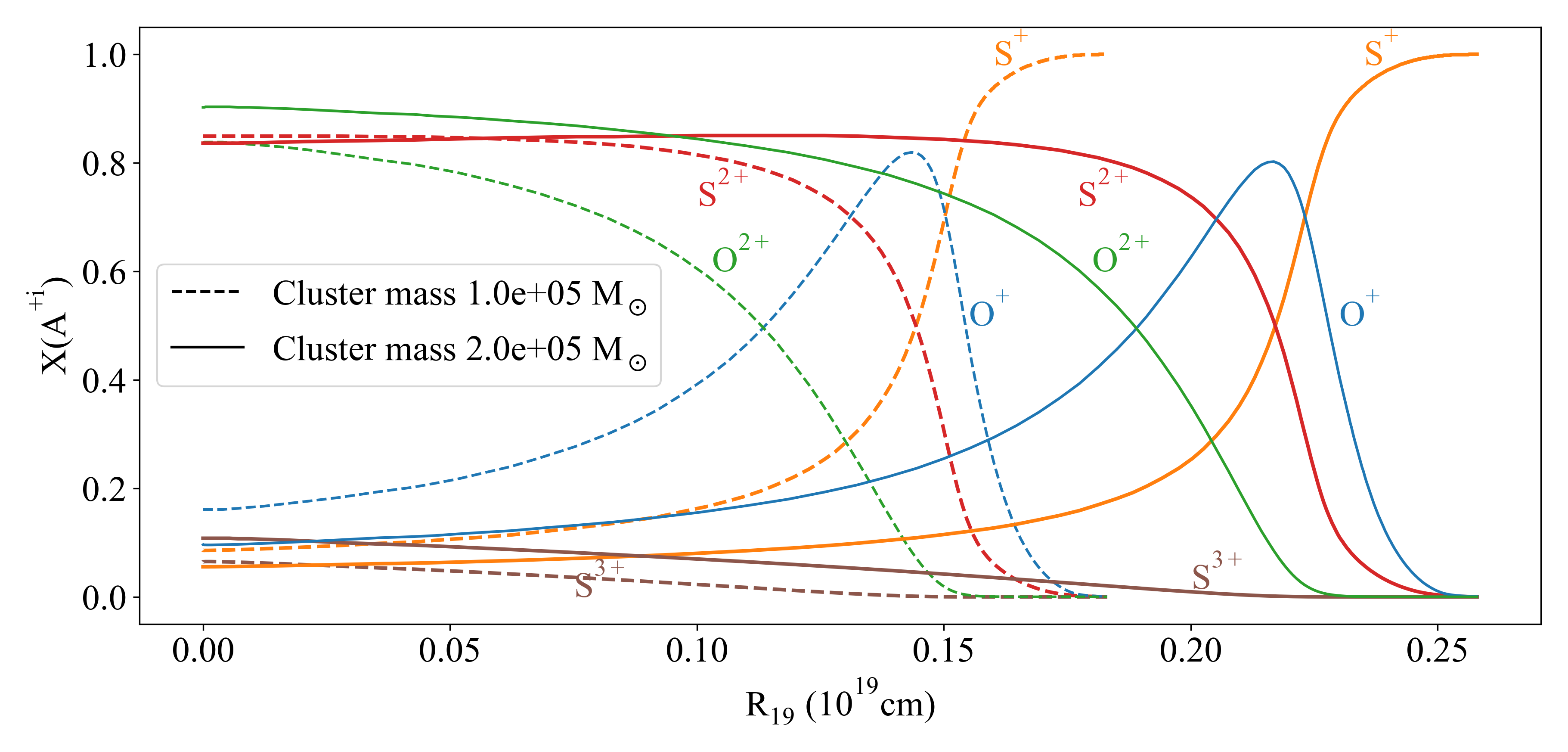}
\par\end{centering}
\caption{\label{fig:IonizationFrac_vs_thickness} Ionization
structure for sulphur and oxygen in two of the PopStar evolutionary
synthesis models described in the text: $log\left(age\right)=5.48$,
$Z_{gas}=0.004$, $log\left(Z_{\star}\right)=-2.1$ and constant $n_{e}=100\,cm^{3}$.
Dashed lines represent a cluster mass of $10^{5}\,M_{\odot}$,
solid lines correspond to $2\cdot10^{5}\,M_{\odot}$.}
\end{figure*}
\begin{figure}
\includegraphics[width=0.5\textwidth]{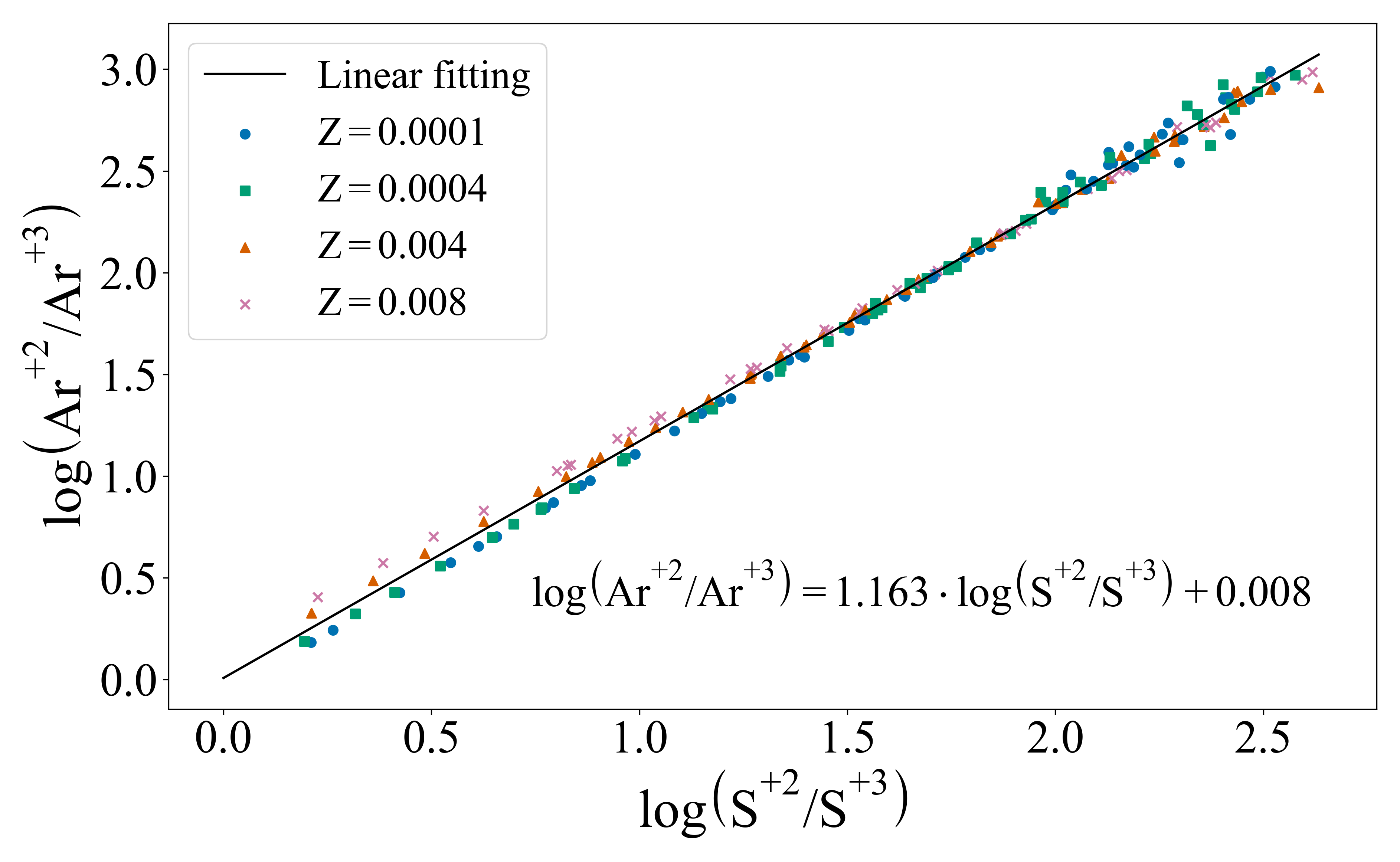}


\caption{\label{fig:Argon-sulphur_Grid} Relation between the ratio of twice
to thrice ionized argon and sulphur abundances for the 
PopStar models grid described in the text.}
\end{figure}
$n_{e}\left(S^{+}\right)$ is commonly assumed in the literature to
characterize HII regions whose densities are below the critical value
for collisional deexcitation ($\sim$ 10$^{4}$). Regarding the electron
temperature, we assume $T_{e}\left(S^{+}\right)\approx T_{e}\left(S^{2+}\right)=T_{e}\left[SIII\right]$.
Justification for this model can be found in the ionization structure
diagrams, such as those introduced by \cite{garnett1992}. Fig.\ref{fig:IonizationFrac_vs_thickness}
was generated using the data from the evolutionary synthesis models developed
by \cite{martin-manjon2010} within the PopStar project. The radiation
spectrum along with the ionizing photons, inner radius and gas metallicity
were used as the input conditions in Cloudy 17.01 \citep{ferland2017}.
This enabled us to simulate a wide range of starforming
regions. The model grid is characterized by $log\left(age\right)=5.0-10.0$,
$Z_{gas}=0.0004-0.05$, $Mass=12000-200000\,M_{\odot}$, $log\left(Z_{\star}\right)=-2.1$
and $n_{e}=100\,cm^{-3}$. 

Fig. \ref{fig:IonizationFrac_vs_thickness} displays
the ionization structure from two of the models: oxygen and sulphur
for $10^{5}M_{\odot}$ and $2\cdot10^{5}M_{\odot}$ cluster masses.
A sharp transition between $S^{+}$ and $S^{2+}$ ionization states
can be seen. Both ions are observed in the transition zone
and a common electron temperature may be assumed. Henceforth, this
temperature is referred to as $T_{low}$ and the sulphur abundance
is calculated as:
\begin{eqnarray}
n_{e}\left(S^{+}\right)=n_{e}\left[SII\right] & = & f\left(\frac{I\left(\left[SII\right]6716\AA\right)}{I\left(\left[SII\right]6731\AA\right)}\right)\nonumber \\
T_{low}=T_{e}\left[SIII\right] & = & f\left(\frac{I\left(\left[SIII\right]6312\AA\right)}{I\left(\left[SIII\right]9069\AA+\left[SIII\right]9532\AA\right)}\right)\nonumber \\
\frac{S^{+}}{H^{+}} & = & f\left(\sum_{i}I_{i}(\left[SII\right]),\,T_{low},\,n_{e}\left[SII\right]\right)\nonumber \\
\frac{S^{2+}}{H^{+}} & = & f\left(\sum_{i}I_{i}(\left[SIII\right]),\,T_{low},\,n_{e}\left[SII\right]\right)\label{eq:sulphur_steps}
\end{eqnarray}

Due to the low ionization potential for sulphur, the neutral S fraction
can be neglected. However, this may not be true for
the $S^{3+}$ population. A direct $S^{3+}$ determination would require
the flux from the infrared $\left[SIV\right]10.51\mu m$ line. As
an alternative we propose an individualized ICF (Ionization Correction
Factor) based on argon. The $Ar^{+2}/Ar^{+3}$ vs. $S^{2+}/S^{3+}$
relation is shown in Fig. \ref{fig:Argon-sulphur_Grid} for the models
grid previously described. A linear relation for both elements at
low gas metallicity is seen in the figure. A linear fitting gives:
\begin{align}
log\left(\frac{Ar^{2+}}{Ar^{3+}}\right) & =a\cdot log\left(\frac{S^{2+}}{S^{3+}}\right)+b\nonumber \\
a =1.162\pm0.006;\,b = 0.05\pm0.01\label{eq:ICF_sulfur}
\end{align}

This ICF is suitable for optical observations as transitions from
the $Ar^{+3}$ ion are observable in the optical range. The argon
ionic ratio is calculated as:

{\footnotesize{}
\begin{eqnarray}
\frac{Ar^{2+}}{H^{+}} & = & f\left(\sum_{i}I_{i}(\left[ArIII\right]),\,T_{low},\,n_{e}\left[SII\right]\right)\nonumber \\
\frac{Ar^{3+}}{H^{+}} & = & f\left(\sum_{i}I_{i}(\left[ArIV\right]),\,T_{high},\,n_{e}\left[SII\right]\right)
\end{eqnarray}
}
where $T_{high}$ is the temperature for the high ionization region
and the $T\left(\left[OIII\right]\right)$ temperature is used to
characterize it. Its calculation is detailed in section \ref{subsec:Oxygen abundance}.
The total sulphur abundance is defined as: 
\begin{equation}
\frac{S}{H}=\left(\frac{S^{+}}{H^{+}}+\frac{S^{2+}}{H^{+}}\right)\cdot ICF\left(S^{3+}\right)
\end{equation}
We set $ICF\left(S^{3+}\right)=1$ for objects where the $Ar^{3+}$
lines are not observed, and consequently assumed $\nicefrac{S^{3+}}{H^{+}}\approx0$.

\subsubsection{\label{subsec:Oxygen abundance}Oxygen}

It is commonly accepted that the $O^{2+}$ ion
characterizes the temperature of the highly ionized gas in these metal
poor star forming galaxies. This assumes $T_{e}\left(O^{2+}\right)\approx T_{e}\left[OIII\right]$.
Contrary to sulphur discussed above, $T_{e}\left(O^{+}\right)\nsim T_{e}\left(O^{2+}\right)$.
Indeed, the $O^{+}$ ions are situated in a lower ionization region
with a lower mean temperature. For this zone it is possible to adopt
$T_{e}\left[OII\right]\approx T_{e}\left[SIII\right]$ \cite{haegele2006,haegele2008}.
To calculate the ionic abundances, the $O^{2+}$ fraction can be accurately
measured from the $\left[OIII\right]\lambda\lambda4959\AA,5007\AA$
lines. Ideally the $O^{+}$ abundance should be measured from the
strong $\left[OII\right]\lambda\lambda3726\AA,3729\AA$ doublet. Unfortunately,
these lines are not available for objects observed with our configuration
I (see Table \ref{tab:Observational-log}). For these objects we use
the $\left[OII\right]\lambda\lambda7319\AA,7330\AA$ weaker auroral
lines doublet. Although these lines have been used in the literature
to calculate the $O^{+}$abundance \cite[see][]{kniazev2003,pilyugin2006,sanchez-almeida2016}, they have also been reported to increase the dispersion on physical parameters such as the $T_{e}\left[OII\right]$ as described by \citet{robertc.kennicutt2003}. These authors argue that one phenomenon
contributing to these deviations is the recombination contribution
to the lines. To account for this feature, we used the correction
provided by \cite{liu2000}:

\begin{equation}
\frac{I\left(\left[OII\right]7319\AA+\left[OII\right]7330\AA\right)}{I\left(H\beta\right)}=9.36\cdot t^{0.44}\times\frac{O^{2+}}{H^{+}}\label{eq:liu_7320_recombination}
\end{equation}

Therefore:

\begin{eqnarray}
T_{high}=T_{e}\left[OIII\right] & = & f\left(\frac{I\left(\left[OIII\right]4363\AA\right)}{I\left(\left[SOIII\right]4959\AA+\left[OIII\right]5007\AA\right)}\right)\nonumber \\
\frac{O^{+}}{H^{+}} & = & f\left(\sum_{i}I_{i}(\left[OII\right]),\,T_{low},\,n_{e}\left[SII\right]\right)\nonumber \\
\frac{O^{2+}}{H^{+}} & = & f\left(\sum_{i}I_{i}(\left[OIII\right]),\,T_{high},\,n_{e}\left[SII\right]\right)
\end{eqnarray}
and the total oxygen abundance is calculated as:

\begin{equation}
\frac{O}{H}=\left(\frac{O^{+}}{H^{+}}+\frac{O^{2+}}{H^{+}}\right)
\end{equation}

The values for the ionic and total O abundance are given in Tables
\ref{tab:Ionic-abundances-2} and \ref{tab:Elemental-abundances-and-Y}
respectively.

\subsubsection{\label{subsec:Nitrogen abundance}Nitrogen:}

In the optical wavelength range, the nitrogen abundance is normaly
derived from the $\left[NII\right]\lambda\lambda6548\AA,6583\AA$
lines. Our spectral resolution is enough to discriminate both lines
from $H\alpha$. Nevertheless, as explained in Sec. \ref{subsec:Emission-Spectrum-1},
some objects required a fitting of four Gaussians to remove the contribution
from a broad $H\alpha$ component.

A strong linear relation appears to exist between the $T_{e}\left[NII\right]$
and $T_{e}\left[SIII\right]$ values \citep{croxall2015} which validates
the practice of adopting $T_{e}\left(\left[NII\right]\right)\approx T_{e}\left(\left[SIII\right]\right)$ 
\citep{perez-montero2003}. For a given atomic data set, the $N^{+}$
abundance is defined as: 
\begin{equation}
\frac{N^{+}}{H^{+}}=f\left(\sum_{i}I_{i}(\left[NII\right]),\,T_{low},\,n_{e}\left[SII\right]\right)
\end{equation}

Using the standard approach by \cite{peimbert1969} which states that
$NI/OI\approx NII/OII$, the nitrogen abundance can be calculated
as: 
\begin{align}
\frac{N}{H} & =\frac{N}{O}\cdot\frac{O}{H}\nonumber \\
 & =\frac{\nicefrac{NII}{HII}}{\nicefrac{OII}{HII}}\cdot\frac{O}{H}\label{eq:NO_relation}
\end{align}

Our derived ionic and total nitrogen abundances are
given in Tables \ref{tab:Ionic-abundances-2} and \ref{tab:Elemental-abundances-and-Y}
respectively.

\section{Results and Discussion}

Table 2 includes emission lines data for two galaxies
observed using ISIS-I and ISIS-II set ups respectively\footnote
{The complete table for all the objects is published in electronic
form}. The table lists the equivalent width, the observed flux and
the reddening-corrected intensity (where $H\beta=1000$). These entries
include the correction from the underlying stellar continuum fitting.

For $^{4}$He to be considered a ``baryometer\textquotedbl{}\footnote{This term, coined by David Schramm, reflects the dependence
of primordial light element abundances on the cosmic density of baryons
and the baryon to photon ratio: $\eta$} its primordial abundance ($Y_{P}$) needs to be determined with
high accuracy taking into account the many possible error sources
\citep[e.g.][]{davidson1985} in particular the systematic errors \citep[e.g.][]{peimbert2007}.
The uncertainties can be separated into 5 groups: data acquisition
and reduction, quantifying the spectra components, ionic abundances
calculation procedures, ionization correction factors and total abundances
determination, and finally, errors associated with the extrapolation
to zero metals.

In what follows we will use our data to derive $Y_{P}$ as the extrapolation
to zero metals of the relation between He and metal abundances, using
as metallicity indicator sulphur and compare the results to those
obtained using the classical approach of oxygen and nitrogen as metallicity
indicators. We will also discuss the impact on the $Y_{P}$ value
of the different sources of error.

\subsection{Data and reduction uncertainties}

Observational data from the near-infrared (NIR) is essential for accurate
sulphur abundance determination. The $\left[SIII\right]\lambda\lambda9069,9532\AA$
 lines are prominent features
in this wavelength range. We have discussed in section \ref{Observations}
the difficulties that arise from observing in this region and how
to circumvent them. 

Since our current reduction pipeline does propagate the pixel flux
uncertainty, theoretical ionic emissivity ratios may be compared with
the observed ionic flux ratios as a qualitative check. The first and
second column in Table \ref{tab-sample-properties-HII-galaxy}
show the observed ratios
$F\left(\left[OIII\right]\lambda5007\AA\right)/F\left(\lambda4959\AA\right)$
and $F\left(\left[SIII\right]\lambda9532\AA\right)/F\left(\lambda9069\AA\right)$
for our sample. The magnitudes are colour coded green, orange
and red to represent deviations from the theoretical value by below
5\%, below 10\% and above 10\% respectively. A very good agreement
for the oxygen ratio in the blue spectrograph arm can be noticed.
Regarding the sulphur ratio in the red arm, a close to theoretical
ratio is obtained when telluric corrections could be performed.

These results suggest a flux calibration within 5\%
 which is the error assumed for the standard
stars flux calibration curves.

\subsection{Flux uncertainty}

Flux uncertainties can have an impact on the physical parameters,
e.g.~a 10\% uncertainty yields a temperature difference in $T\left[SIII\right]$
of around $500\,K$. This translates into an abundance uncertainty of
around $0.15\,dex$ \cite[see][]{robertc.kennicutt2003,haegele2008,dorsjr.2016}.
Fortunately, the random nature
of this error minimizes its effect on the linear regression $Y_{P}$
vs. S/H as long as its magnitude remains small.

\subsection{Nebular continuum uncertainty}

Our sample selection criterion of $EW\left(H\alpha\right)>200\,\AA$
guarantees an ionizing cluster younger than $10\,Myr$ \citep[see][]{dottori1981,leitherer1999}.
The youngest burst in our sample would be that of $SHOC220$
with $EW\left(H\alpha\right)=1310\AA$. In such young bursts, the
luminosity is dominated by the emission. However, to guarantee
the most accurate abundance derivation from the weak helium lines,
the underlying stellar absorption needs to be quantified for which 
previous subtraction of the nebular continuum is required.

The nebular continuum can be calculated from first principles, as
described in section \ref{subsec:Nebular-continuum-section}. This
calculation, however, requires the input of some characteristic electron
temperature and density, as well as flux. These parameters can be
inferred from the data when it includes the Balmer and/or Paschen
jumps. In the former case, the continuum temperature $\left(T_{Bac}\right)$
can be related to the jump flux intensity $\left(BJ\right)$ and the
$H_{11}\lambda3770\AA$ Balmer line, following \cite{liu2001}: 
\begin{equation}
T_{Bac}=368\cdot\left(1+0.259\cdot y^{+}+3.409\cdot y^{2+}\right)\left(\frac{BJ}{H_{11}}\right)\label{eq:balmerJump}
\end{equation}
where the Balmer jump flux
is calculated assuming linear continuum regions before and after the
discontinuity. In this work (see \ref{subsec:Nebular-continuum-section})
the nebular continuum intensity was calibrated using a Zanstra-like
approach, eq. \ref{eq:zanstra_calibration}, using a hydrogen recombination
line. We used $H\alpha$ due to its relatively low contamination
by stellar absorption. The de-reddened spectrum of SHOC579 is shown
in figure \ref{fig:SHOC579 Spectra-components-comparison}, where
we can distinguish the small Balmer jump and the $H_{20}\lambda3683\AA$
line. Using eq. \ref{eq:balmerJump}, $T_{Bac}=13200\pm3000\,K$ in agreement with the values estimated from the oxygen and sulfur emission line diagnostics, although this measurement has a larger uncertainty. This was
also the conclusion reached by \citet{garcia-rojas2012} for planetary
nebulae: The large number of emission lines in the proximity of the
hydrogen discontinuities leaves very few points to perform the linear
fits. The situation is worse for the Paschen jump due to the telluric contamination on the continuum. An optimum solution
would consist of fitting simultaneously the nebular and stellar continua
to the observed spectrum. This approach has been applied successfully
by \citet{gomes2017} in their FADO package. Fortunately, for the electron temperatures and densities commonly
encountered in HII galaxies, the  featureless nebular continuum is almost independent
of these parameters \citep{zhang2004}. Consequently, for this region, the nebular continuum shape can be obtained assuming $T_{e}=10000\,K$ and $n_{e}=100.0$. This is also the approach employed by FADO for input spectra which don't include the hydrogen discontinuities.

\subsection{Uncertainties in the underlying population subtraction}

\begin{figure*}
\includegraphics[width=1\textwidth]{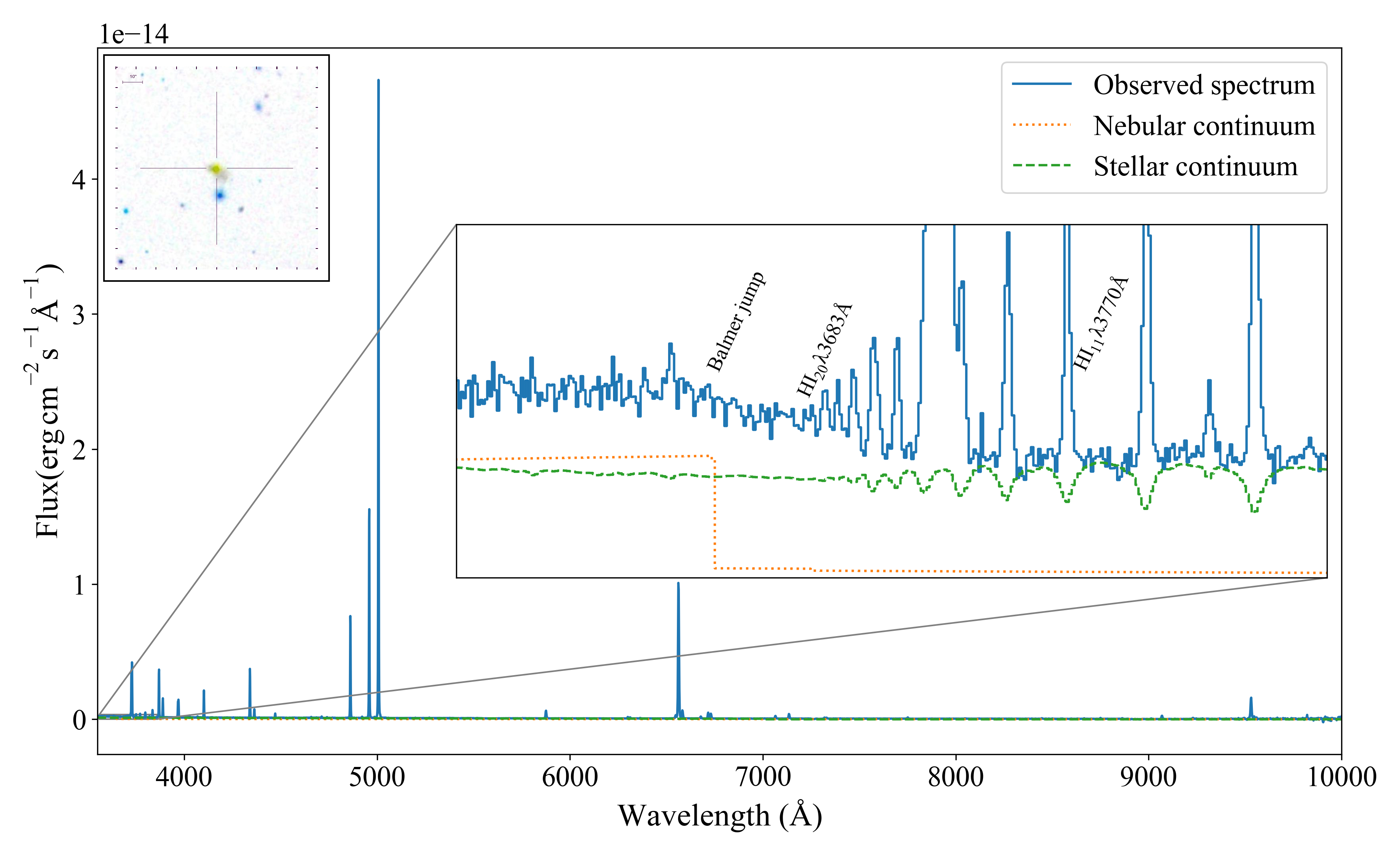}

\caption{\label{fig:SHOC579 Spectra-components-comparison}Spectral components
for SHOC579. The object image belongs to the SDSS
database.}
\end{figure*}

\cite{rosales-ortegaf.f.2006} illustrated the impact of including a spectral synthesis analysis on the helium abundance calculation. The absorption on the recombination lines can be checked in Fig. \ref{fig:SHOC579 Spectra-components-comparison}, which displays the fitted stellar continuum for SHOC579 using STARLIGHT. Our SSP library is particularly tailored for young and low metallicity populations (as described in section \ref{subsec:Stellar-Continuum}). In these galaxies, as it can be seen on the figure, the absorption features may be absent on the observed spectrum. Consequently, it is challenging for spectra synthesis codes to fit a stellar continuum. Within our SSPs library wavelength coverage, our spectra do not include any significant absorption features. To avoid overestimations on the stellar velocity dispersion, our fittings are constrained to a maximum calculated from the mean velocity dispersion of the CaII triplet lines. If these were not observed, the $\left[OIII\right]\lambda5007\AA$ velocity dispersion was adopted $(20<\sigma_{\left[OIII\right]\lambda5007\AA}<80\,km/s)$.

{\footnotesize{}}
\begin{figure*}
{\footnotesize{}\includegraphics[width=0.5\textwidth]{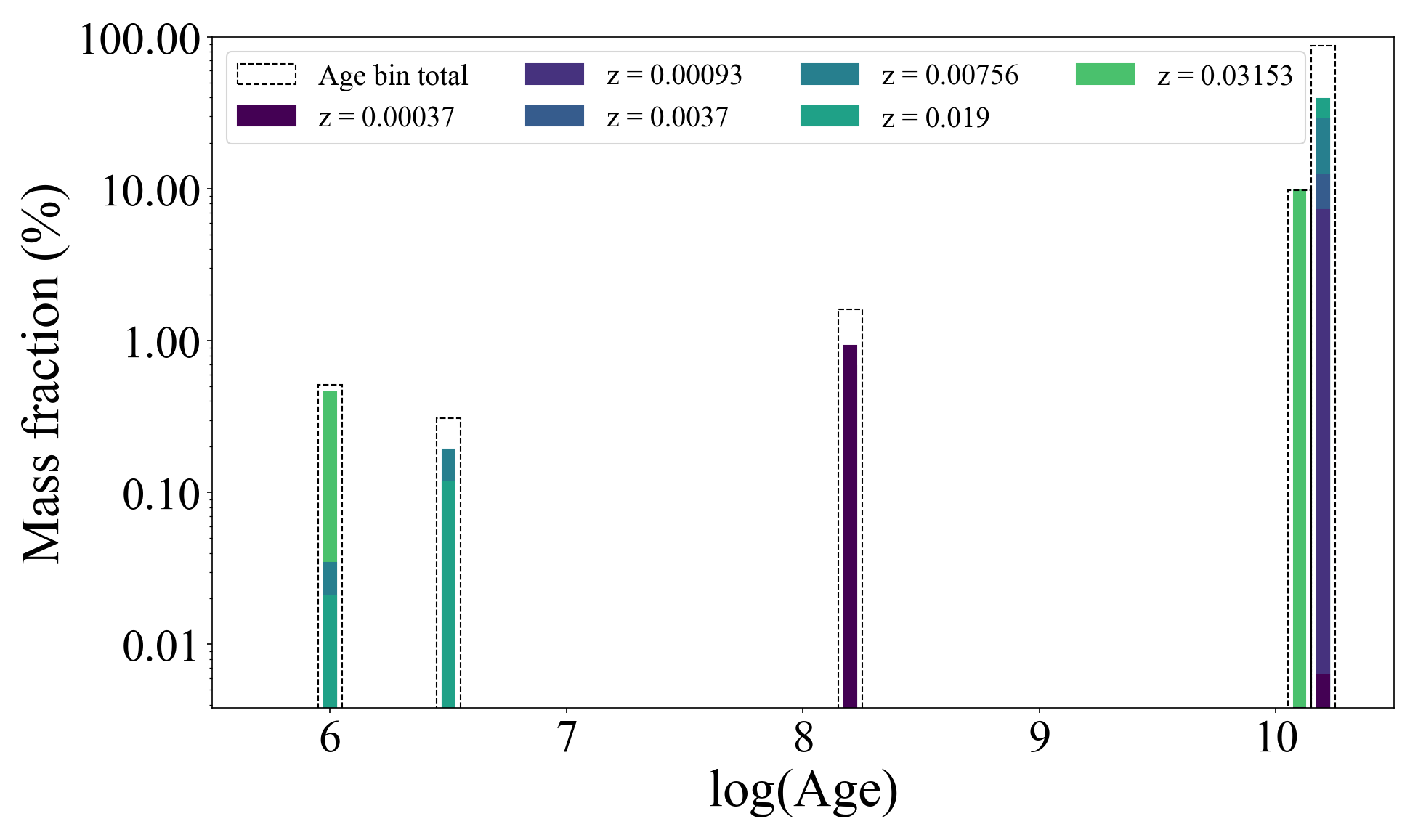}\includegraphics[width=0.5\textwidth]{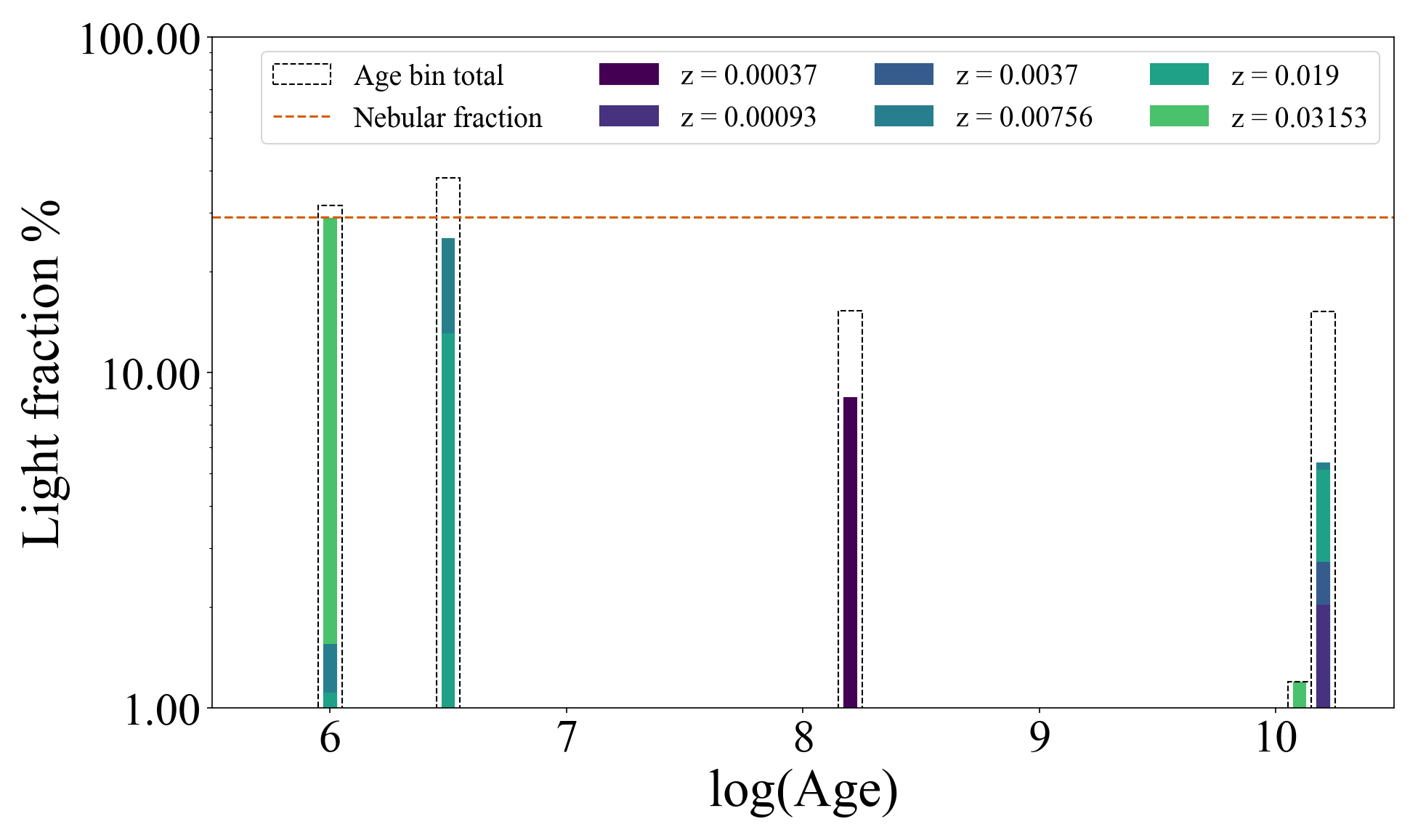}}{\footnotesize \par}

{\footnotesize{}\caption{\label{fig:Starlight_Histograms}Mass (left) and light (right) weighted
population histograms for SHOC579 from STARLIGHT. The bar colours
represent the stellar population metallicity, while the dotted line
represents the total contribution from several metallicities. The
dotted horizontal line represents the light from the nebular continuum
as a percentage of the total stellar continuum.}
}{\footnotesize \par}
\end{figure*}
 The histograms in Fig. \ref{fig:Starlight_Histograms} display the
stellar population output from the STARLIGHT fit to SHOC579. The left
and right panels  correspond to the mass and light weighted
population distribution respectively versus the age bin. 
The
metallicities are colour coded in the $0.00037<Z<0.03153$ interval.
The implications from these results are as  expected: the older
populations dominate the mass, while the younger one dominates the
light. A metallicity gradient may be noted with older stars displaying
a lower metal content. Additional information which can be derived
from this analysis includes the current stellar mass. For SHOC579
$log\left(M_{T}\right)=8.05$  which implies that the
host galaxy should be relatively evolved, although we cannot say anything
from the observed spectrum. Finally, to compare
the nebular and stellar continuum luminosity, we integrate the total
flux from each continuum for our wavelength range. This is displayed
on the right hand side histogram in Fig. \ref{fig:Starlight_Histograms}
as a dashed horizontal line that quantifies the nebular-to-stellar
luminosity percentage. For SHOC579, the nebular luminosity is almost
20\% of the stellar luminosity and is of the same order as the brightest
population. This illustrates the impact of the gas luminosity whenever
we seek to quantify the youngest populations on  bursts of star formation.

While the input spectra on these STARLIGHT fittings
have been corrected from reddening, the program also computes
a uniform extinction of the stellar populations. For most objects
this extinction is very low or zero suggesting that the extinction
in the ionized gas is similar to that of the immersed stellar population.
However, this is not the case for SHOC579 for which $A_{V;\,stellar}=0.43$,
while the extinction predicted from the Balmer decrement is $A_{V;\,gas}=0.061\pm0.002$
assuming a \cite{gordon2003} extinction law with $R_{V}=3.4$. The impact
on the reddening correction is further discussed in the next section.

The goodness of the STARLIGHT fitting depends largely 
on the available amount of unmasked spectrum and on the intensity of the absorption
features.  Configuration II covers
almost $1000\AA$ more than  Configuration I. In the latter
case, the unmasked continuum is practically featureless: no sizable
absorptions are detected even on the helium lines. In these cases,
the SSP synthesis result consists in just a single young stellar population.
Quantifying the accuracy and precision of these fits remains a challenge
for a SSP algorithm \citep{cidfernandes2013}. Still, even with a
poor stellar population fit, the current analysis should not be less accurate 
than results using other methods
for quantifying the underlying stellar population in previous $Y_{P}$
determinations. We have already mentioned the approach by \cite{rosales-ortegaf.f.2006} followed by \cite{peimbert2007} who computed the helium
abundance taking $EW_{ab}\approx2\AA$ from \cite{gonzalez-delgado2005}
models. In the multidimensional analysis by \cite{aver2015}, they
consider two constant values for
the underlying absorption in each object: one for the hydrogen lines
and another one for the helium lines. Finally, \cite{izotov2014}, uses for
all the HII regions an underlying absorption for each helium line
from the models of \cite{gonzalez-delgado2005} and the Starburst99
high resolution spectra \citep{leitherer1999}.

The solution towards a better stellar continuum fit relies on increasing the
wavelength coverage while preserving the temporal and
spectral resolution. Currently, the simple stellar libraries by \citep{bruzual2003}, which reach $9500\,\AA$, are being updated with younger stellar populations\footnote{Soon to be available, Aida Wofford and Gustavo Bruzual, private communication.}. Consequently, it will be possible to fit the CaII triplet $(\lambda8498\AA,8542\AA,8662\AA)$ in observations such as these ones. This will provide an anchor to the older stellar population.

\subsection{T$_{e}$, n$_{e}$ and ionic abundances}

At this point the emission line fluxes can be used
to calculate $T_{e}$ and $n_{e}$. The extinction still needs to
be calculated, though. A canonical approach consists in comparing the
theoretical ratio of the Balmer lines with the observed flux ratios
along with a reddening law. In HII regions, however, there is some
degeneracy in this approach since the observed recombination lines
could also be affected by underlying stellar absorption. The solution
we chose is to apply the following scheme:
\begin{enumerate}
\item An initial reddening correction is obtained from
the Balmer decrement assuming $T_{e}=10000\,K$ and $n_{e}=100\,cm^{-3}$;
the theoretical ratios are calculated using PyNeb. We use \cite{gordon2003}
reddening law as appropriate for young starforming bursts with $R_{V}=3.4$. 
\item The ionic abundances are calculated. 
\item The nebular contribution is calculated using the
physical parameters determined from the previous step and the dereddened
$H\alpha$ emission.
\item The nebular contribution is subtracted from the
dereddened object spectrum and the stellar continuum is fitted. 
\item The nebular and stellar continua are removed from
the spectrum and the corrected emission lines are measured. 
\item A new reddening coefficient is calculated. The previous
steps are repeated using the measurements from this iteration as initial
values.
\item From the second iteration we obtain the final line
intensities and from them the element abundances as discussed in section
\ref{sec:Chemical-analysis}, which we will use in a regression to
estimate $Y_{P}$ as we will discuss in the following section.
\end{enumerate}
Since both the extinction and underlying stellar
absorption to the recombination lines increase towards shorter wavelengths
the reddening coefficient decreases in the second iteration. In one
object, it decreased to almost a half of its original value. However,
since most of these objects show low extinction this change did not
produce a great impact on the ionic abundances. As mentioned in the
previous section, the stellar bases employed on the stellar population
synthesis did not reach beyond $7000\,\AA$. Consequently, none of
the  Paschen  lines could be included on the reddening
calculation as they were not corrected from stellar absorption. Still,
even if this would had been the case, the Paschen lines within our wavelength
range $\left(P_{20}\,to\,P_{9}\right)$ have a
low signal to noise ratio due to the telluric contamination. Consequently,
their weight on the $c\left(H\text{\ensuremath{\beta}}\right)$ linear
fit is very low compared to the one from the Balmer $H\alpha,$ $H\beta$,
$H\gamma$ and $H\delta$ lines. As an additional precaution, to avoid
flux calibration mismatch between the spectrograph arms, the lines
in the blue arm were normalized with respect to $H\beta$, while for those
in the red arm we used the most intense hydrogen recombination line
available depending on the configuration: $H\alpha$, $P_{8}$ or
$P_{9}$. Afterwards, these lines were multiplied by the theoretical
emissivity ratio $\left(\epsilon_{\lambda,red\,arm}/\epsilon_{H\beta}\right)$
so that the lines in both arms could be compared.

A direct abundance estimate, such as the one described in section
\ref{sec:Chemical-analysis}, requires measurement of both temperature
and density sensitive line ratios. For our data we can determine $T_{e}\left[OIII\right]$ and $T_{e}\left[SIII\right]$
for both configurations. These values are given in Table 
\ref{tab-sample-properties-HII-galaxy}.
$T_{e}\left[OIII\right]$ has been exclusively used in the helium
abundance determination. This is due to the characteristics of the
objects involved: low metallicity and high ionization factors. This
guarantees  strong lines, hence good precision. In
contrast the $T_{e}\left[SIII\right]$ can be measured in ionized
gas with metallicity  up to solar. 
This is an important asset in order to determine better the slope of the relation and therefore to extend the linear regression
for determining $Y_{P}$  (see the following subsection). On the 
other hand, stronger lines imply smaller uncertainty for the gradient
\citep[][CHAOS survey]{berg2015}. Berg and collaborators use $T\left[SIII\right]$,
$T\left[NII\right]$ and the theoretical relation by \cite{garnett1992}
to estimate the high ionization $T_{e}\left[OIII\right]$.This was shown in Fig. \ref{fig:Temperature-comparison-O-S}. The orthogonal
linear regression from our data agrees with the estimates in the literature.
The dotted line represents the estimate from \cite{garnett1992} while
the dashed lines are the photo-ionization models by \cite{perez-montero2005}
and \cite{perez-montero2014}. The different gradients in these two linear models can be explained by the different atomic data sources for the $S^{2+}$ collisional strengths.
It should be noticed that all lines overlap in the $10,000-15,000K$
region. This is also the region, where $T_{e}\left[SIII\right]\backsim T_{e}\left[OIII\right]$
and where most of our data falls. The well known extremely low metallicity
HII galaxy IZw18 lies in the outskirts of this relation displaying
a higher than expected $T_{e}\left[SIII\right]$. Its chemical analysis,
however, is still under debate. In a recent paper by \cite{kehrig2016},
this galaxy was analyzed using integral field spectroscopy to compare
the emission from both NW and SE galaxy knots. The NW region is larger,
has a higher electron temperature, lower metallicity and a stronger
continuum. In contrast, the smaller SE region has slightly higher
metallicity and lower electron temperature. Their
estimates: $T_{e}\left[OIII\right]_{NW}=23000\pm700\,K$ and $T_{e}\left[OIII\right]_{SW}=19600\pm600\,K$
agree within errors with our values shown in Table \ref{tab-sample-properties-HII-galaxy}.
Given the linear relation between both temperatures for our sample
objects, as discussed in section 3.5, we adopted the following compromise: the most precise temperature (either $T_{e}\left[OIII\right]$ or $T_{e}\left[SIII\right]$)
is used to characterize the galaxy. For example,
if a low redshift object has a large error for the $\left[OIII\right]\lambda4363\AA$
line because it lies very close to the spectrum edge, the high ionization
temperature will be calculated using $T_{e}\left[SIII\right]$ in
eq. \ref{eq:TSIII_TOIII_relation}. This will minimize the dispersion
on the regression hence improving the extrapolation for determining
$Y_{P}$, as will be discussed in the following sections.

\begin{table*}
\caption{\label{tab-sample-properties-HII-galaxy}Electron temperature and
density. A superscript on the object name represents
the element used for the temperature. In these cases the missing low
or high ionization temperature was derived from eq. \ref{eq:TSIII_TOIII_relation}.
The colour coding in columns (2) and (3) represents the departure
from the theoretical ratio: green below 5\%, yellow below 10\% and
red above 10\% }

\begin{tabu}{lccccc}%
\hline%
HII Galaxy&$\frac{[OIII]\lambda5007\AA}{[OIII]\lambda4959\AA}$&$\frac{[SIII]\lambda9531\AA}{[SIII]\lambda9069\AA}$&$n_{e}[SII](cm^{-3})$&$T_{e}[SIII](K)$&$T_{e}[OIII](K)$\\%
\hline%
FTDTR-1$^{S}$&\textcolor{ForestGreen}{2.99$\pm$0.065}&\textcolor{ForestGreen}{2.39$\pm$0.037}&49.9$\pm$21&14700$\pm$710&17900$\pm$910\\%
IZw18$^{S}$&\textcolor{ForestGreen}{3.08$\pm$0.079}&\textcolor{ForestGreen}{2.56$\pm$0.067}&49.7$\pm$22&21700$\pm$1800&18300$\pm$900\\%
MRK36-A1$^{}$&\textcolor{ForestGreen}{3.01$\pm$0.078}&\textcolor{ForestGreen}{2.39$\pm$0.020}&50.0$\pm$22&15100$\pm$670&14900$\pm$850\\%
MRK36-A2$^{}$&\textcolor{ForestGreen}{3.03$\pm$0.096}&\textcolor{ForestGreen}{2.59$\pm$0.014}&88.3$\pm$31&12900$\pm$290&15800$\pm$430\\%
MRK475$^{}$&\textcolor{ForestGreen}{3.05$\pm$0.091}&\textcolor{YellowOrange}{2.70$\pm$0.013}&50.7$\pm$21&13400$\pm$340&14300$\pm$410\\%
FTDTR-2$^{O}$&\textcolor{ForestGreen}{3.11$\pm$0.083}&\textcolor{ForestGreen}{2.36$\pm$0.058}&50.3$\pm$22&14100$\pm$1700&13900$\pm$760\\%
IZw70$^{S}$&\textcolor{ForestGreen}{2.99$\pm$0.089}&\textcolor{ForestGreen}{2.54$\pm$0.018}&50.8$\pm$22&10800$\pm$450&12900$\pm$710\\%
MRK689$^{S}$&\textcolor{ForestGreen}{3.04$\pm$0.077}&\textcolor{ForestGreen}{2.45$\pm$0.015}&51.5$\pm$22&10200$\pm$360&11100$\pm$850\\%
MRK67$^{S}$&\textcolor{ForestGreen}{3.05$\pm$0.097}&\textcolor{ForestGreen}{2.51$\pm$0.019}&81.5$\pm$38&12600$\pm$450&14800$\pm$610\\%
FTDTR-3$^{}$&\textcolor{ForestGreen}{3.03$\pm$0.071}&\textcolor{ForestGreen}{2.37$\pm$0.011}&49.2$\pm$21&12800$\pm$420&12900$\pm$360\\%
SHOC022$^{}$&\textcolor{ForestGreen}{3.11$\pm$0.034}&\textcolor{ForestGreen}{2.42$\pm$0.020}&49.5$\pm$22&11400$\pm$580&12200$\pm$620\\%
FTDTR-4$^{}$&\textcolor{ForestGreen}{3.09$\pm$0.065}&\textcolor{ForestGreen}{2.41$\pm$0.018}&49.0$\pm$22&12200$\pm$580&15000$\pm$330\\%
SHOC220$^{O}$&\textcolor{ForestGreen}{2.88$\pm$0.0021}&\textcolor{ForestGreen}{2.36$\pm$0.059}&128$\pm$81&17500$\pm$1200&19500$\pm$400\\%
FTDTR-5$^{O}$&\textcolor{ForestGreen}{3.04$\pm$0.065}&\textcolor{ForestGreen}{2.47$\pm$0.073}&165$\pm$110&14400$\pm$2200&13100$\pm$1100\\%
FTDTR-6$^{}$&\textcolor{ForestGreen}{3.04$\pm$0.069}&\textcolor{YellowOrange}{2.60$\pm$0.0086}&49.9$\pm$22&12100$\pm$440&12800$\pm$400\\%
FTDTR-7$^{}$&\textcolor{ForestGreen}{3.05$\pm$0.057}&\textcolor{ForestGreen}{2.50$\pm$0.0085}&132$\pm$20&13300$\pm$300&13800$\pm$200\\%
MRK627$^{S}$&\textcolor{ForestGreen}{3.04$\pm$0.056}&\textcolor{ForestGreen}{2.49$\pm$0.019}&50.6$\pm$22&10300$\pm$460&11300$\pm$770\\%
SHOC592$^{O}$&\textcolor{ForestGreen}{3.02$\pm$0.0066}&\textcolor{YellowOrange}{2.60$\pm$0.13}&149$\pm$30&9740$\pm$750&10400$\pm$300\\%
PHL293B$^{}$&\textcolor{ForestGreen}{3.08$\pm$0.076}&\textcolor{YellowOrange}{2.65$\pm$0.080}&112$\pm$31&14000$\pm$500&15800$\pm$230\\%
SHOC588$^{}$&\textcolor{YellowOrange}{3.15$\pm$0.044}&\textcolor{ForestGreen}{2.54$\pm$0.069}&50.3$\pm$22&11900$\pm$390&10500$\pm$190\\%
SHOC036$^{O}$&\textcolor{ForestGreen}{3.04$\pm$0.013}&-&49.9$\pm$21&-&16300$\pm$550\\%
SHOC575$^{O}$&\textcolor{ForestGreen}{2.96$\pm$0.022}&\textcolor{BrickRed}{1.90$\pm$0.033}&125$\pm$43&11700$\pm$580&10700$\pm$340\\%
SHOC579$^{}$&\textcolor{ForestGreen}{3.05$\pm$0.0031}&\textcolor{BrickRed}{5.80$\pm$0.15}&147$\pm$8.0&11600$\pm$160&12400$\pm$50\\%
FTDTR-8$^{O}$&\textcolor{ForestGreen}{3.00$\pm$0.0060}&\textcolor{BrickRed}{2.96$\pm$0.078}&50.2$\pm$22&13400$\pm$2000&15100$\pm$550\\%
SHOC263$^{O}$&\textcolor{ForestGreen}{2.98$\pm$0.0097}&-&227$\pm$83&-&11000$\pm$400\\%
FTDTR-9$^{}$&\textcolor{ForestGreen}{3.09$\pm$0.076}&\textcolor{YellowOrange}{2.67$\pm$0.069}&147$\pm$62&11400$\pm$890&11600$\pm$780\\%
FTDTR-10$^{O}$&\textcolor{ForestGreen}{2.88$\pm$0.0061}&\textcolor{BrickRed}{2.73$\pm$0.12}&666$\pm$200&-&14200$\pm$720\\%
\hline%
\end{tabu}
\end{table*}
The abundances, as described in section \ref{sec:Chemical-analysis},
are determined using the canonical direct method except for two points.
First, the $O^{+}/H^{+}$ determination for many objects depends
on the $\left[OII\right]\lambda\lambda7319\AA,7330\AA$ lines as the
stronger $\left[OII\right]\lambda\lambda3726\AA,3729\AA$ doublet is
not observed with Configuration I. These weaker lines are affected 
by a recombination effect and in some cases also by
sky emission. These effects are treated using the recombination
prediction from eq. \ref{eq:liu_7320_recombination} and the telluric
correction respectively. The possible underlying stellar absorption can not be
properly accounted for as those lines are beyond the wavelength limit
of the used stellar database, as already discussed. It is expected,
however, that this error should be very small given the young age, by selection,
of the sample. 

The second difference is the calculation of the $ICF\left(S^{3+}\right)$
using argon (see section \ref{subsec:sulphur abundance}
and Fig. \ref{fig:Argon-sulphur_Grid}). Even when $S^{+3}$ is
expected to be present in these objects on account of their high excitation,
still it is hard to quantify. Even though data exists for $\left[SIV\right]10.51\mu m$
or $\left[SIII\right]18.71\mu m$ these lines have  also a recombination
component. It is essential to anchor the infrared results to optical
observations by including measurements of a nearby recombination line.
Fig. \ref{fig:-ICF_sulfur_calculated-from} compares the predictions
using our ICFs with available infrared data for $\left[SIV\right]10.51\mu m$
and $\left[SIII\right]18.71\mu m$ in the literature as compiled by \cite{dorsjr.2016}. While the empirical results are biased towards
low temperatures both ICFs cover the same region. The uncertainty
in these ICFs comes both from the error propagation on the ionic abundances,
and from the fit using eq. \ref{eq:ICF_sulfur}. The helium, oxygen
and nitrogen ionic abundances for our sample are shown in Table \ref{tab:Ionic-abundances},
while Table \ref{tab:Ionic-abundances-2} displays their measured sulphur
and argon ionic abundances.

\begin{table*}
\caption{\label{tab:Ionic-abundances}Helium, oxygen and nitrogen ionic abundances.
}

\begin{tabu}{lccccc}%
\hline%
HII Galaxy&$\nicefrac{He^{+}}{H^{+}}$&$\nicefrac{He^{2+}}{H^{+}}$&$12 + log\left(\nicefrac{O^{+}}{H^{+}}\right)$&$12 + log\left(\nicefrac{O^{2+}}{H^{+}}\right)$&$12 + log\left(\nicefrac{N^{+}}{H^{+}}\right)$\\%
\hline%
\href{http://dr12.sdss3.org/spectrumDetail?mjd=53725\&fiber=536\&plateid=2329}{FTDTR-1}&0.0977$\pm$0.0011&0.0012$\pm$0.0002&7.29$\pm$0.10&7.75$\pm$0.053&5.38$\pm$0.050\\%
\href{http://dr12.sdss3.org/spectrumDetail?mjd=51991\&fiber=312\&plateid=556}{IZw18}&0.0768$\pm$0.0028&0.00053$\pm$0.0&6.48$\pm$0.12&6.95$\pm$0.069&4.74$\pm$0.062\\%
\href{http://dr12.sdss3.org/spectrumDetail?plateid=2211\&mjd=53786\&fiber=486}{MRK36-A1}&0.0795$\pm$0.0082&0.0018$\pm$0.0002&7.32$\pm$0.091&7.73$\pm$0.071&5.56$\pm$0.21\\%
\href{http://dr12.sdss3.org/spectrumDetail?plateid=2211\&mjd=53786\&fiber=486}{MRK36-A2}&0.0801$\pm$0.0097&0.0012$\pm$0.0&7.68$\pm$0.051&7.68$\pm$0.038&5.67$\pm$0.33\\%
\href{http://dr12.sdss3.org/spectrumDetail?plateid=1382\&mjd=53115\&fiber=175}{MRK475}&0.0883$\pm$0.0045&0.0017$\pm$0.0001&7.50$\pm$0.059&7.83$\pm$0.043&5.75$\pm$0.25\\%
\href{http://dr12.sdss3.org/spectrumDetail?plateid=976\&mjd=52413\&fiber=600}{FTDTR-2}&0.0756$\pm$0.0033&0.0010$\pm$0.0002&-&7.88$\pm$0.069&5.53$\pm$0.067\\%
\href{http://dr12.sdss3.org/spectrumDetail?plateid=1383\&mjd=53116\&fiber=110}{IZw70}&0.0943$\pm$0.0032&0.00058$\pm$0.0001&8.15$\pm$0.11&7.99$\pm$0.055&6.40$\pm$0.045\\%
\href{http://dr12.sdss3.org/spectrumDetail?plateid=1388\&mjd=53119\&fiber=17}{MRK689}&0.0775$\pm$0.0068&-&8.35$\pm$0.095&8.19$\pm$0.049&6.31$\pm$0.041\\%
\href{http://dr12.sdss3.org/spectrumDetail?plateid=2094\&mjd=53851\&fiber=487}{MRK67}&0.0910$\pm$0.012&0.00096$\pm$0.0&7.79$\pm$0.084&8.00$\pm$0.047&5.88$\pm$0.19\\%
\href{http://dr12.sdss3.org/spectrumDetail?plateid=2233\&mjd=53845\&fiber=371}{FTDTR-3}&0.0847$\pm$0.0030&-&7.47$\pm$0.074&8.12$\pm$0.043&5.61$\pm$0.24\\%
\href{http://dr12.sdss3.org/spectrumDetail?plateid=418\&mjd=51817\&fiber=302}{SHOC022}&0.0838$\pm$0.0018&0.0014$\pm$0.0001&7.97$\pm$0.13&7.97$\pm$0.069&6.24$\pm$0.055\\%
\href{http://dr12.sdss3.org/spectrumDetail?plateid=861\&mjd=52318\&fiber=489}{FTDTR-4}&0.0878$\pm$0.0015&0.00075$\pm$0.0&7.87$\pm$0.11&7.85$\pm$0.028&6.04$\pm$0.049\\%
\href{http://dr12.sdss3.org/spectrumDetail?plateid=550\&mjd=51959\&fiber=92}{SHOC220}&0.0989$\pm$0.0023&0.0018$\pm$0.0001&6.80$\pm$0.043&7.52$\pm$0.024&5.56$\pm$0.075\\%
\href{http://dr12.sdss3.org/spectrumDetail?plateid=942\&mjd=52703\&fiber=612}{FTDTR-5}&0.0779$\pm$0.0042&-&-&8.06$\pm$0.11&-\\%
\href{http://dr12.sdss3.org/spectrumDetail?plateid=456\&mjd=51910\&fiber=76}{FTDTR-6}&0.0869$\pm$0.0019&0.00088$\pm$0.0&7.80$\pm$0.090&8.00$\pm$0.045&5.84$\pm$0.16\\%
\href{http://dr12.sdss3.org/spectrumDetail?plateid=1185\&mjd=52642\&fiber=123}{FTDTR-7}&0.0864$\pm$0.0013&0.00041$\pm$0.0&7.49$\pm$0.050&8.05$\pm$0.020&5.72$\pm$0.022\\%
\href{http://dr12.sdss3.org/spectrumDetail?plateid=934\&mjd=52672\&fiber=369}{MRK627}&0.0845$\pm$0.0040&-&8.30$\pm$0.12&8.12$\pm$0.060&6.54$\pm$0.051\\%
\href{http://dr12.sdss3.org/spectrumDetail?plateid=640\&mjd=52200\&fiber=270}{SHOC592}&0.0962$\pm$0.0080&-&7.74$\pm$0.061&8.18$\pm$0.043&6.59$\pm$0.037\\%
\href{http://dr12.sdss3.org/spectrumDetail?plateid=376\&mjd=52143\&fiber=160}{PHL293B}&0.0766$\pm$0.0024&0.0017$\pm$0.0002&6.73$\pm$0.052&7.70$\pm$0.021&-\\%
\href{http://dr12.sdss3.org/spectrumDetail?plateid=639\&mjd=52146\&fiber=242}{SHOC588}&0.0940$\pm$0.0084&0.00059$\pm$0.0&7.38$\pm$0.055&8.18$\pm$0.028&6.25$\pm$0.034\\%
\href{http://dr12.sdss3.org/spectrumDetail?plateid=394\&mjd=51913\&fiber=472}{SHOC036}&0.102$\pm$0.010&0.0028$\pm$0.0007&6.75$\pm$0.046&7.69$\pm$0.038&5.57$\pm$0.051\\%
\href{http://dr12.sdss3.org/spectrumDetail?plateid=358\&mjd=51818\&fiber=472}{SHOC575}&0.0977$\pm$0.0018&-&7.57$\pm$0.063&8.18$\pm$0.046&6.66$\pm$0.038\\%
\href{http://dr12.sdss3.org/spectrumDetail?plateid=358\&mjd=51818\&fiber=504}{SHOC579}&0.0991$\pm$0.0049&0.00040$\pm$0.0&7.17$\pm$0.023&8.05$\pm$0.0057&6.35$\pm$0.014\\%
\href{http://dr12.sdss3.org/spectrumDetail?plateid=942\&mjd=52703\&fiber=612}{FTDTR-8}&0.0750$\pm$0.0018&-&7.12$\pm$0.11&7.86$\pm$0.045&5.27$\pm$0.096\\%
\href{http://dr12.sdss3.org/spectrumDetail?plateid=556\&mjd=51991\&fiber=224}{SHOC263}&0.0866$\pm$0.0056&-&8.19$\pm$0.11&8.04$\pm$0.057&6.63$\pm$0.053\\%
\href{http://dr12.sdss3.org/spectrumDetail?plateid=501\&mjd=52235\&fiber=602}{FTDTR-9}&0.0890$\pm$0.0032&0.0016$\pm$0.0001&8.00$\pm$0.23&8.05$\pm$0.095&6.27$\pm$0.086\\%
\href{http://dr12.sdss3.org/spectrumDetail?plateid=575\&mjd=52319\&fiber=521}{FTDTR-10}&0.0783$\pm$0.0026&0.0010$\pm$0.0002&7.22$\pm$0.13&7.79$\pm$0.071&5.79$\pm$0.21\\%
\hline%
\end{tabu}
\end{table*}

\begin{table*}
\caption{\label{tab:Ionic-abundances-2}Sulphur and argon ionic abundances.}

\begin{tabu}{lccccc}%
\hline%
HII Galaxy&$12 + log\left(\nicefrac{S^{+}}{H^{+}}\right)$&$12 + log\left(\nicefrac{S^{2+}}{H^{+}}\right)$&$ICF\left(S^{3+}\right)$&$12 + log\left(\nicefrac{Ar^{2+}}{H^{+}}\right)$&$12 + log\left(\nicefrac{Ar^{3+}}{H^{+}}\right)$\\%
\hline%
\href{http://dr12.sdss3.org/spectrumDetail?mjd=53725\&fiber=536\&plateid=2329}{FTDTR-1}&5.15$\pm$0.041&5.83$\pm$0.036&1.95$\pm$0.078&5.16$\pm$0.043&5.18$\pm$0.062\\%
\href{http://dr12.sdss3.org/spectrumDetail?mjd=51991\&fiber=312\&plateid=556}{IZw18}&4.62$\pm$0.050&5.08$\pm$0.048&1.37$\pm$0.046&4.55$\pm$0.056&4.16$\pm$0.090\\%
\href{http://dr12.sdss3.org/spectrumDetail?plateid=2211\&mjd=53786\&fiber=486}{MRK36-A1}&5.31$\pm$0.038&5.97$\pm$0.035&1.26$\pm$0.045&5.40$\pm$0.037&4.78$\pm$0.080\\%
\href{http://dr12.sdss3.org/spectrumDetail?plateid=2211\&mjd=53786\&fiber=486}{MRK36-A2}&5.43$\pm$0.024&6.16$\pm$0.021&1.22$\pm$0.018&5.53$\pm$0.023&4.79$\pm$0.037\\%
\href{http://dr12.sdss3.org/spectrumDetail?plateid=1382\&mjd=53115\&fiber=175}{MRK475}&5.45$\pm$0.027&6.23$\pm$0.024&1.20$\pm$0.019&5.64$\pm$0.026&4.85$\pm$0.043\\%
\href{http://dr12.sdss3.org/spectrumDetail?plateid=976\&mjd=52413\&fiber=600}{FTDTR-2}&5.28$\pm$0.054&5.99$\pm$0.046&1.50$\pm$0.082&5.38$\pm$0.058&5.07$\pm$0.10\\%
\href{http://dr12.sdss3.org/spectrumDetail?plateid=1383\&mjd=53116\&fiber=110}{IZw70}&5.91$\pm$0.043&6.40$\pm$0.037&-&5.80$\pm$0.042&-\\%
\href{http://dr12.sdss3.org/spectrumDetail?plateid=1388\&mjd=53119\&fiber=17}{MRK689}&5.93$\pm$0.039&6.56$\pm$0.033&-&5.84$\pm$0.038&-\\%
\href{http://dr12.sdss3.org/spectrumDetail?plateid=2094\&mjd=53851\&fiber=487}{MRK67}&5.54$\pm$0.037&6.21$\pm$0.033&1.32$\pm$0.041&5.68$\pm$0.037&5.15$\pm$0.056\\%
\href{http://dr12.sdss3.org/spectrumDetail?plateid=2233\&mjd=53845\&fiber=371}{FTDTR-3}&5.23$\pm$0.032&6.16$\pm$0.029&1.88$\pm$0.084&5.59$\pm$0.031&5.53$\pm$0.040\\%
\href{http://dr12.sdss3.org/spectrumDetail?plateid=418\&mjd=51817\&fiber=302}{SHOC022}&5.77$\pm$0.053&6.42$\pm$0.045&1.28$\pm$0.054&5.74$\pm$0.051&5.14$\pm$0.087\\%
\href{http://dr12.sdss3.org/spectrumDetail?plateid=861\&mjd=52318\&fiber=489}{FTDTR-4}&5.64$\pm$0.046&6.24$\pm$0.040&1.19$\pm$0.023&5.64$\pm$0.045&4.87$\pm$0.038\\%
\href{http://dr12.sdss3.org/spectrumDetail?plateid=550\&mjd=51959\&fiber=92}{SHOC220}&4.73$\pm$0.024&5.58$\pm$0.017&1.79$\pm$0.062&5.02$\pm$0.030&4.93$\pm$0.028\\%
\href{http://dr12.sdss3.org/spectrumDetail?plateid=942\&mjd=52703\&fiber=612}{FTDTR-5}&5.30$\pm$0.090&6.05$\pm$0.076&1.79$\pm$0.25&5.47$\pm$0.093&5.37$\pm$0.13\\%
\href{http://dr12.sdss3.org/spectrumDetail?plateid=456\&mjd=51910\&fiber=76}{FTDTR-6}&5.48$\pm$0.037&6.36$\pm$0.033&1.26$\pm$0.032&5.72$\pm$0.036&5.06$\pm$0.050\\%
\href{http://dr12.sdss3.org/spectrumDetail?plateid=1185\&mjd=52642\&fiber=123}{FTDTR-7}&5.27$\pm$0.020&6.09$\pm$0.017&1.69$\pm$0.040&5.50$\pm$0.019&5.34$\pm$0.020\\%
\href{http://dr12.sdss3.org/spectrumDetail?plateid=934\&mjd=52672\&fiber=369}{MRK627}&5.99$\pm$0.049&6.55$\pm$0.041&-&5.89$\pm$0.049&-\\%
\href{http://dr12.sdss3.org/spectrumDetail?plateid=640\&mjd=52200\&fiber=270}{SHOC592}&6.07$\pm$0.036&6.69$\pm$0.031&1.19$\pm$0.077&5.89$\pm$0.036&5.10$\pm$0.23\\%
\href{http://dr12.sdss3.org/spectrumDetail?plateid=376\&mjd=52143\&fiber=160}{PHL293B}&5.12$\pm$0.031&5.95$\pm$0.031&1.57$\pm$0.054&5.35$\pm$0.031&5.09$\pm$0.036\\%
\href{http://dr12.sdss3.org/spectrumDetail?plateid=639\&mjd=52146\&fiber=242}{SHOC588}&5.73$\pm$0.032&6.35$\pm$0.029&1.22$\pm$0.041&5.80$\pm$0.031&5.09$\pm$0.089\\%
\href{http://dr12.sdss3.org/spectrumDetail?plateid=394\&mjd=51913\&fiber=472}{SHOC036}&5.22$\pm$0.033&5.91$\pm$0.031&1.77$\pm$0.16&5.25$\pm$0.049&5.16$\pm$0.099\\%
\href{http://dr12.sdss3.org/spectrumDetail?plateid=358\&mjd=51818\&fiber=472}{SHOC575}&5.74$\pm$0.037&6.48$\pm$0.031&-&5.95$\pm$0.036&-\\%
\href{http://dr12.sdss3.org/spectrumDetail?plateid=358\&mjd=51818\&fiber=504}{SHOC579}&5.54$\pm$0.014&6.36$\pm$0.015&1.31$\pm$0.016&5.76$\pm$0.013&5.19$\pm$0.020\\%
\href{http://dr12.sdss3.org/spectrumDetail?plateid=942\&mjd=52703\&fiber=612}{FTDTR-8}&5.14$\pm$0.040&5.89$\pm$0.032&1.86$\pm$0.082&5.42$\pm$0.042&5.37$\pm$0.059\\%
\href{http://dr12.sdss3.org/spectrumDetail?plateid=556\&mjd=51991\&fiber=224}{SHOC263}&6.00$\pm$0.046&6.60$\pm$0.049&-&5.87$\pm$0.044&-\\%
\href{http://dr12.sdss3.org/spectrumDetail?plateid=501\&mjd=52235\&fiber=602}{FTDTR-9}&5.74$\pm$0.083&6.52$\pm$0.069&-&5.73$\pm$0.081&-\\%
\href{http://dr12.sdss3.org/spectrumDetail?plateid=575\&mjd=52319\&fiber=521}{FTDTR-10}&5.52$\pm$0.057&6.18$\pm$0.048&-&5.46$\pm$0.056&-\\%
\hline%
\end{tabu}
\end{table*}

\begin{table*}
\caption{\label{tab:Elemental-abundances-and-Y}Element abundances and helium
mass fractions using either oxygen $\left(Y_{\nicefrac{O}{H}}\right)$
or sulphur $\left(Y_{\nicefrac{S}{H}}\right)$ in
eq. \ref{eq:Y_calculation}.}

{\footnotesize{}\begin{tabu}{lcccccc}%
\hline%
HII Galaxy&$\nicefrac{He}{H}$&$Y_{\left(\nicefrac{O}{H}\right)}$&$Y_{\left(\nicefrac{S}{H}\right)}$&$12 + log\left(\nicefrac{O}{H}\right)$&$12 + log\left(\nicefrac{N}{H}\right)$&$12 + log\left(\nicefrac{S}{H}\right)$\\%
\hline%
FTDTR-1\textsuperscript{O, N, S}&0.0977$\pm$0.0011&0.283$\pm$0.0023&0.283$\pm$0.0024&7.88$\pm$0.064&5.97$\pm$0.039&6.20$\pm$0.047\\%
IZw18\textsuperscript{O, N, S}&0.0768$\pm$0.0028&0.236$\pm$0.0065&0.236$\pm$0.0064&7.08$\pm$0.082&5.34$\pm$0.038&5.34$\pm$0.055\\%
MRK36-A1\textsuperscript{O, N, S}&0.0795$\pm$0.0082&0.245$\pm$0.019&0.244$\pm$0.019&7.87$\pm$0.059&6.11$\pm$0.21&6.15$\pm$0.034\\%
MRK36-A2\textsuperscript{O, N, S}&0.0801$\pm$0.0097&0.244$\pm$0.022&0.246$\pm$0.023&7.98$\pm$0.033&5.97$\pm$0.33&6.32$\pm$0.020\\%
MRK475\textsuperscript{O, N, S}&0.0883$\pm$0.0045&0.264$\pm$0.0096&0.265$\pm$0.0098&8.00$\pm$0.037&6.25$\pm$0.25&6.37$\pm$0.023\\%
FTDTR-2\textsuperscript{S}&0.0756$\pm$0.0033&-&0.235$\pm$0.0080&-&-&6.24$\pm$0.057\\%
IZw70\textsuperscript{O, N, S}&0.0943$\pm$0.0032&0.274$\pm$0.0067&0.275$\pm$0.0068&8.38$\pm$0.086&6.64$\pm$0.028&6.52$\pm$0.038\\%
MRK689\textsuperscript{N, S}&0.0775$\pm$0.0068&0.235$\pm$0.016&0.236$\pm$0.016&8.57$\pm$0.076&6.54$\pm$0.025&6.66$\pm$0.034\\%
MRK67\textsuperscript{O, N, S}&0.0910$\pm$0.012&0.267$\pm$0.026&0.269$\pm$0.025&8.21$\pm$0.044&6.31$\pm$0.19&6.42$\pm$0.030\\%
FTDTR-3\textsuperscript{O, N, S}&0.0847$\pm$0.0030&0.252$\pm$0.0067&0.253$\pm$0.0069&8.20$\pm$0.038&6.35$\pm$0.24&6.48$\pm$0.025\\%
SHOC022\textsuperscript{O, N, S}&0.0838$\pm$0.0018&0.253$\pm$0.0041&0.254$\pm$0.0040&8.27$\pm$0.073&6.55$\pm$0.040&6.61$\pm$0.040\\%
FTDTR-4\textsuperscript{O, N, S}&0.0878$\pm$0.0015&0.261$\pm$0.0033&0.262$\pm$0.0032&8.16$\pm$0.059&6.34$\pm$0.021&6.42$\pm$0.036\\%
SHOC220\textsuperscript{}&0.0989$\pm$0.0023&0.287$\pm$0.0046&0.287$\pm$0.0047&7.60$\pm$0.022&6.36$\pm$0.079&5.89$\pm$0.019\\%
FTDTR-5\textsuperscript{S}&0.0779$\pm$0.0042&-&0.238$\pm$0.010&-&-&6.37$\pm$0.073\\%
FTDTR-6\textsuperscript{O, N, S}&0.0869$\pm$0.0019&0.259$\pm$0.0043&0.260$\pm$0.0042&8.22$\pm$0.045&6.26$\pm$0.16&6.51$\pm$0.029\\%
FTDTR-7\textsuperscript{O, N, S}&0.0864$\pm$0.0013&0.257$\pm$0.0028&0.258$\pm$0.0028&8.16$\pm$0.019&6.39$\pm$0.027&6.38$\pm$0.014\\%
MRK627\textsuperscript{O, N, S}&0.0845$\pm$0.0040&0.251$\pm$0.0089&0.252$\pm$0.0085&8.52$\pm$0.096&6.77$\pm$0.032&6.66$\pm$0.042\\%
SHOC592\textsuperscript{}&0.0957$\pm$0.0079&0.275$\pm$0.016&0.276$\pm$0.017&8.32$\pm$0.048&7.13$\pm$0.025&6.86$\pm$0.044\\%
PHL293B\textsuperscript{O, S}&0.0766$\pm$0.0024&0.238$\pm$0.0056&0.238$\pm$0.0057&7.74$\pm$0.020&-&6.21$\pm$0.025\\%
SHOC588\textsuperscript{}&0.0940$\pm$0.0084&0.273$\pm$0.018&0.273$\pm$0.018&8.25$\pm$0.025&7.12$\pm$0.029&6.53$\pm$0.028\\%
SHOC036\textsuperscript{}&0.102$\pm$0.010&0.293$\pm$0.020&0.292$\pm$0.020&7.74$\pm$0.038&6.55$\pm$0.049&6.24$\pm$0.051\\%
SHOC575\textsuperscript{}&0.0977$\pm$0.0018&0.280$\pm$0.0038&0.281$\pm$0.0038&8.28$\pm$0.049&7.37$\pm$0.026&6.55$\pm$0.032\\%
SHOC579\textsuperscript{}&0.0991$\pm$0.0049&0.284$\pm$0.0099&0.284$\pm$0.0098&8.11$\pm$0.0057&7.28$\pm$0.0082&6.54$\pm$0.013\\%
FTDTR-8\textsuperscript{O, N, S}&0.0750$\pm$0.0018&0.230$\pm$0.0041&0.231$\pm$0.0042&7.93$\pm$0.050&6.09$\pm$0.11&6.23$\pm$0.041\\%
SHOC263\textsuperscript{O, N, S}&0.0866$\pm$0.0056&0.256$\pm$0.012&0.257$\pm$0.013&8.42$\pm$0.085&6.87$\pm$0.041&6.69$\pm$0.046\\%
FTDTR-9\textsuperscript{O, N, S}&0.0890$\pm$0.0032&0.265$\pm$0.0068&0.266$\pm$0.0064&8.33$\pm$0.12&6.62$\pm$0.069&6.58$\pm$0.071\\%
FTDTR-10\textsuperscript{O, N, S}&0.0783$\pm$0.0026&0.240$\pm$0.0060&0.240$\pm$0.0058&7.89$\pm$0.080&6.48$\pm$0.21&6.27$\pm$0.048\\%
\hline%
\end{tabu}}{\footnotesize \par}
\raggedright{}The superscript on the names indicates the Y, Z regressions,
where the object is included.
\end{table*}

\subsection{$Y_{P}$ regressions}

After the pioneering \citet{peimbert1974} paper, $Y_{P}$ has been
calculated as the extrapolation to zero metals (oxygen) of the He
abundance by mass. Fig. \ref{fig:Primordial-linear-regressions} shows
$\Delta Y\,vs\,\Delta Z$ using as metallicity indicator (Z) oxygen,
nitrogen or sulphur. The abundances are given in Table \ref{tab:Elemental-abundances-and-Y}
where, as a superscript to the object name on Column (1), we indicate
which objects were used for each $Y_{P}$ determination (using either
O, N, or S as metal indicator). The helium abundance is given in
column (2); the helium mass fractions, calculated using either
oxygen or sulphur are columns (3) and (4). Columns (4), (5)
and (6) show the oxygen, nitrogen and sulphur abundances  respectively. 

\begin{figure}
\begin{centering}
\includegraphics[width=0.5\textwidth]{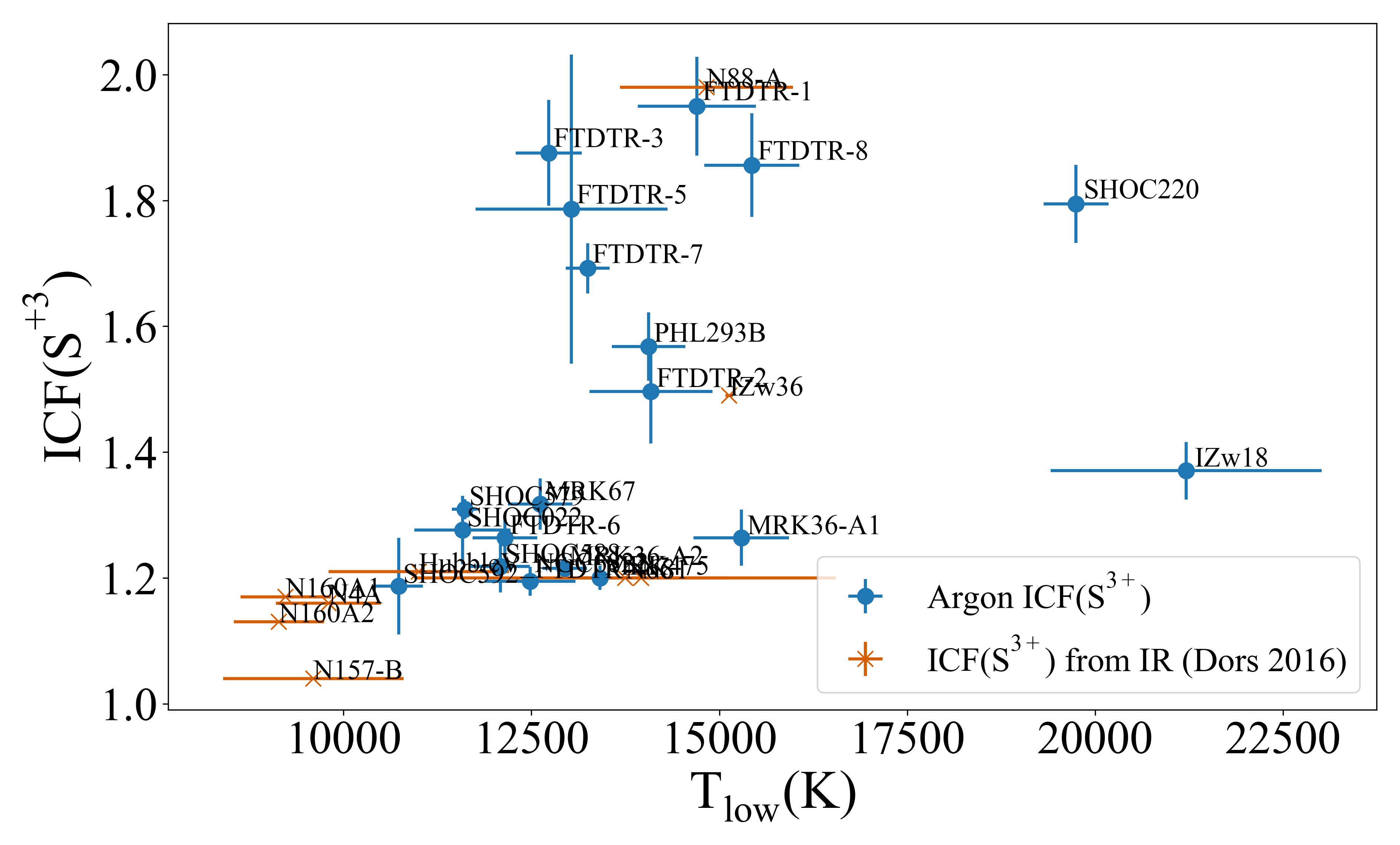}
\par\end{centering}
\caption{\label{fig:-ICF_sulfur_calculated-from} $ICF\left(S^{3+}\right)$
vs. the low ionization temperature. ICFs for our objects (blue points)
were calculated from the Ar lines following the correlation in eq. \ref{eq:ICF_sulfur}.
The red symbols are objects from Dors et al. (2016) with 
[SIII] and 
[SIV] infrared observations.}
\end{figure}
The $\Delta Y\,vs\,\Delta Z$ gradient has been explained by chemical
evolution models \cite[e.g.][]{carigi2011}. A strong Y-N correlation
was also found by \citet{pagel1992} and confirmed by \citet{rosales-ortegaf.f.2006}. This was an unexpected result given the primary/secondary nature of
N: the C\textendash N\textendash O cycle from low mass stars and the
$\alpha$ process in massive stars. In contrast, the oxygen and 
sulphur nucleosynthesis can only be originated by the second mechanism.
It has been claimed that  the agreement between the Y-O and Y-N regressions
is expected under the premises: no serious pollution from Wolf-Rayet stars
and low metallicities $\left(\nicefrac{N}{H}<6.6\cdot10^{-6}\right)$.
The He-S regression proposed in this paper is also fairly linear for
the metallicity range of our sample. Despite the predicted intrinsic
linearity in the chemical evolution between helium and these heavy
elements, some objects seem to be clear outliers in the relation.
In what follows we propose reasons that could explain these outliers.

\begin{figure*}
\begin{centering}
\includegraphics[height=0.3\textheight]{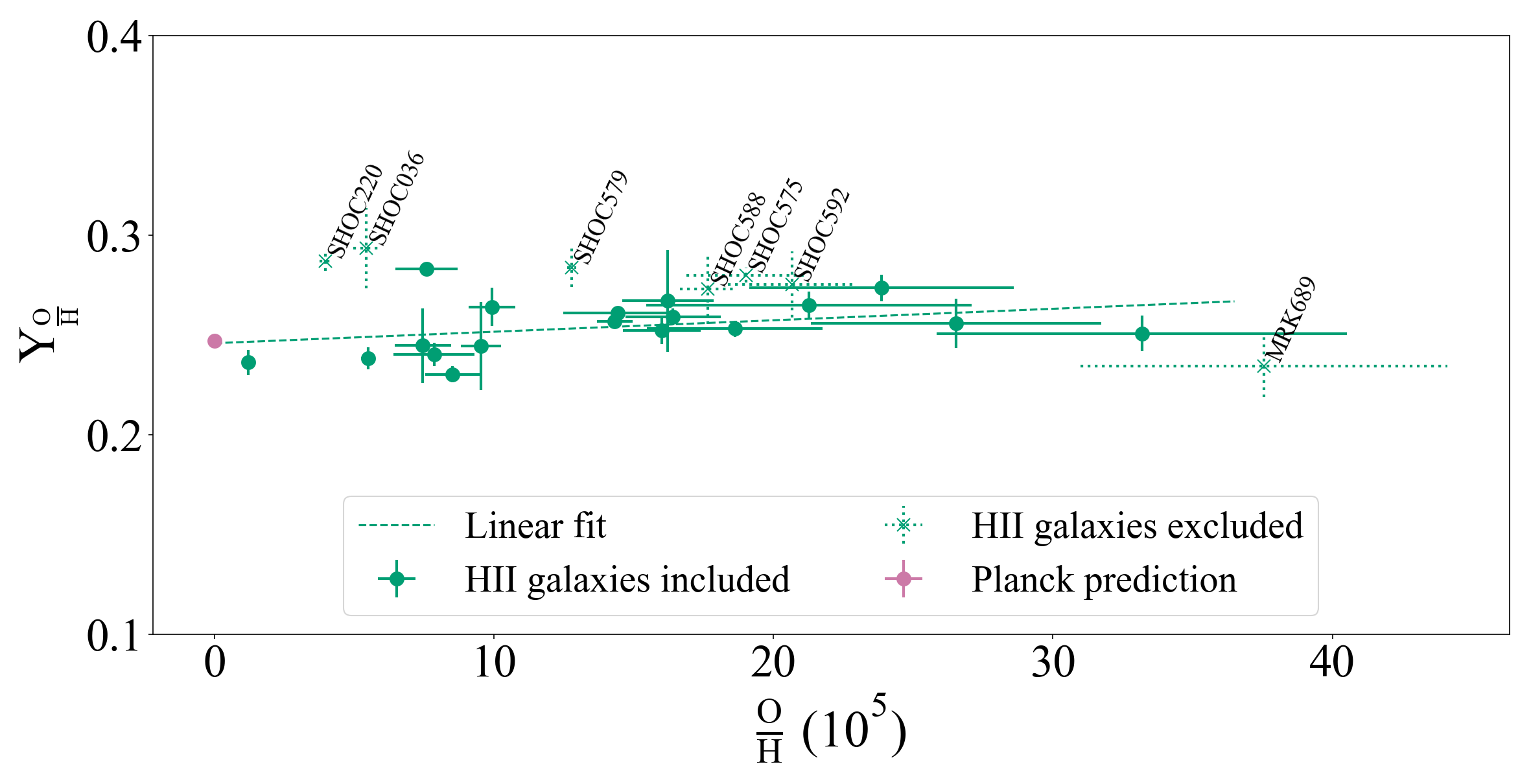}
\par\end{centering}
\begin{centering}
\includegraphics[height=0.3\textheight]{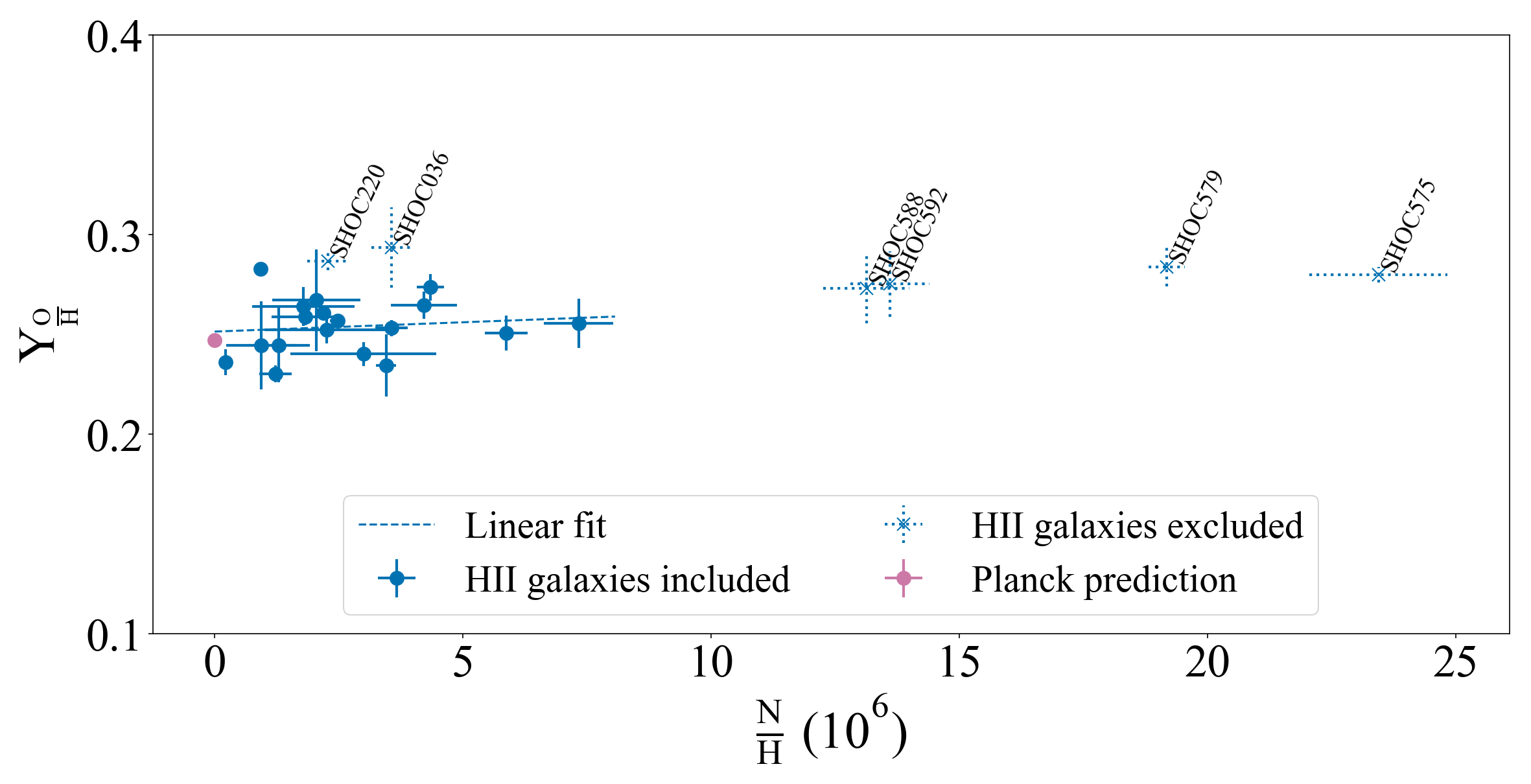}
\par\end{centering}
\begin{centering}
\includegraphics[height=0.3\textheight]{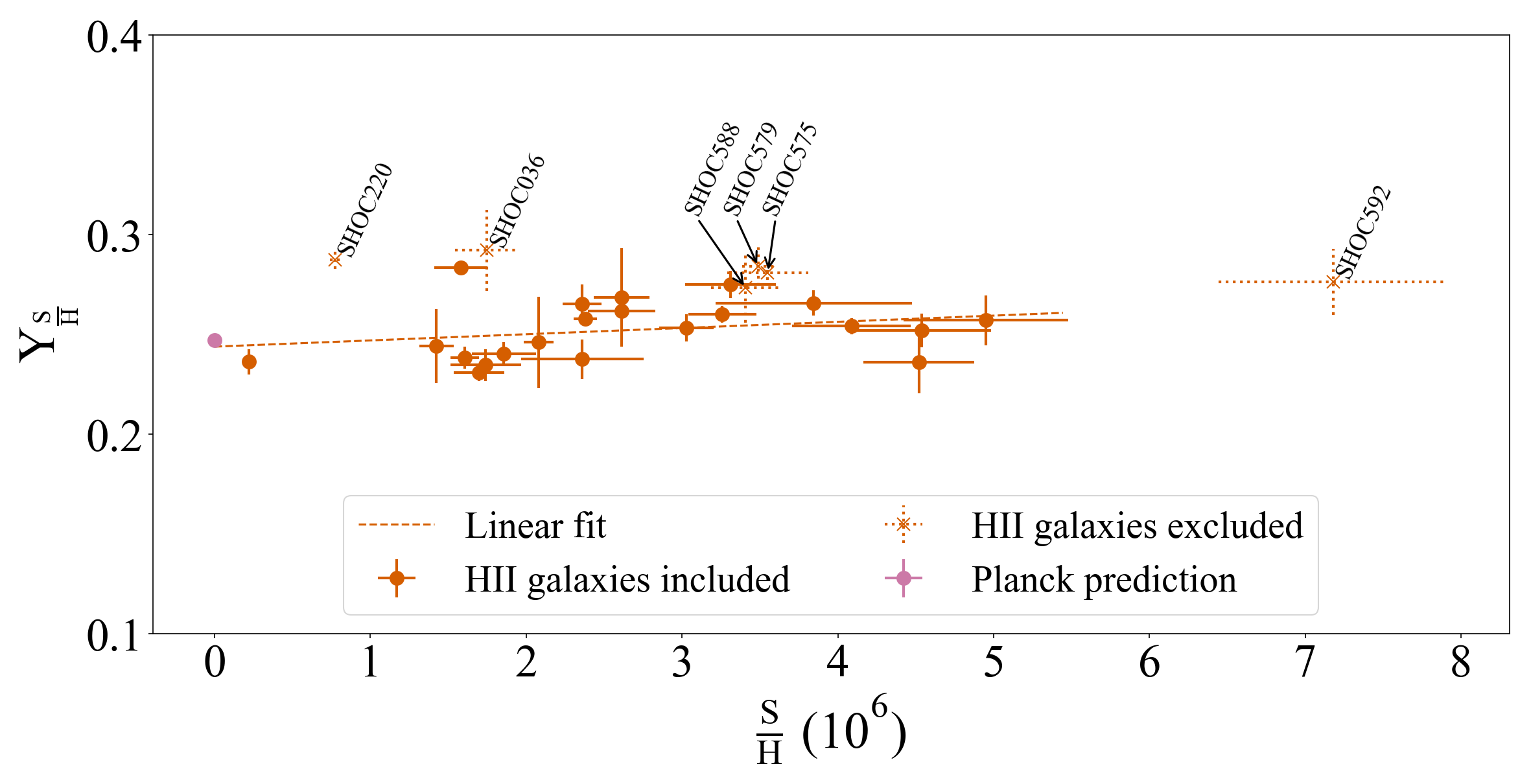}
\par\end{centering}
\caption{\label{fig:Primordial-linear-regressions}Primordial helium linear regressions
using oxygen, nitrogen and sulphur as metallicity tracer.}
\end{figure*}

The top panel of Fig. \ref{fig:Primordial-linear-regressions} shows
the regression of helium vs oxygen abundances. Indicated are the objects
that were excluded from the fit. The galaxies with the higher O/H
abundance appear to display a lower than expected Y abundance. All
these objects were observed with Configuration I, which does not reach
the $\left[OII\right]\lambda\lambda3726\AA,3729\AA$ doublet. Consequently,
in these low ionization objects, where $\nicefrac{O^{+}}{H^{+}}>\nicefrac{O^{++}}{H^{+}}$
there is a greater uncertainty in the $O^{+}$ as its value depends
on weak lines. The temperatures deduced for them are also the lowest
of the sample. Among them is $MRK689$, having the largest error bars
both in $O$ and in $Y$ and for which we derived the highest $O/H$
value, so this object weights heavily  in the linear fit for the regression.
For all these reasons, we did not include it in the fit. Below we  discuss
other objects that were excluded from the final regression.

The middle panel of Fig. \ref{fig:Primordial-linear-regressions} shows
the helium vs. nitrogen regression. Two objects, $SHOC220$ and $PHL293$,
show indications of complex kinematics. $PHL293$ has been extensively
analyzed by \cite{terlevich2014}. Its spectrum shows two very broad
($\sim1000\,km/s$ and $\sim4000\,km/s$) $H\alpha$ components which
the authors argue are generated by a young and dense expanding supershell
or a stationary cooling wind, both driven by the young cluster wind.
The galaxy $SHOC220$ displays similarly intense wide $H\alpha$ components.
Since these broad components, overlap the $\left[NII\right]\lambda\lambda6548\AA,6583\AA$
lines, these galaxies were excluded from the $Y-N$
regression.

Finally, the bottom panel shows the Y vs. S regression where all objects
are included even when for several objects the telluric correction
for the $\left[SIII\right]\lambda\lambda9069,9532\AA$ lines could
not be performed. This can be appreciated on the observed ratios between
these lines in Table \ref{tab-sample-properties-HII-galaxy}. For
these objects, the theoretical relation between both lines was used 
based on the line of the doublet that falls on the least perturbed
sky region. $SHOC036$ and $SHOC263$ were observed during adverse
conditions and none of the $\left[SIII\right]$ lines were detected.

\begin{figure*}
\begin{centering}
\includegraphics[height=0.3\textheight]{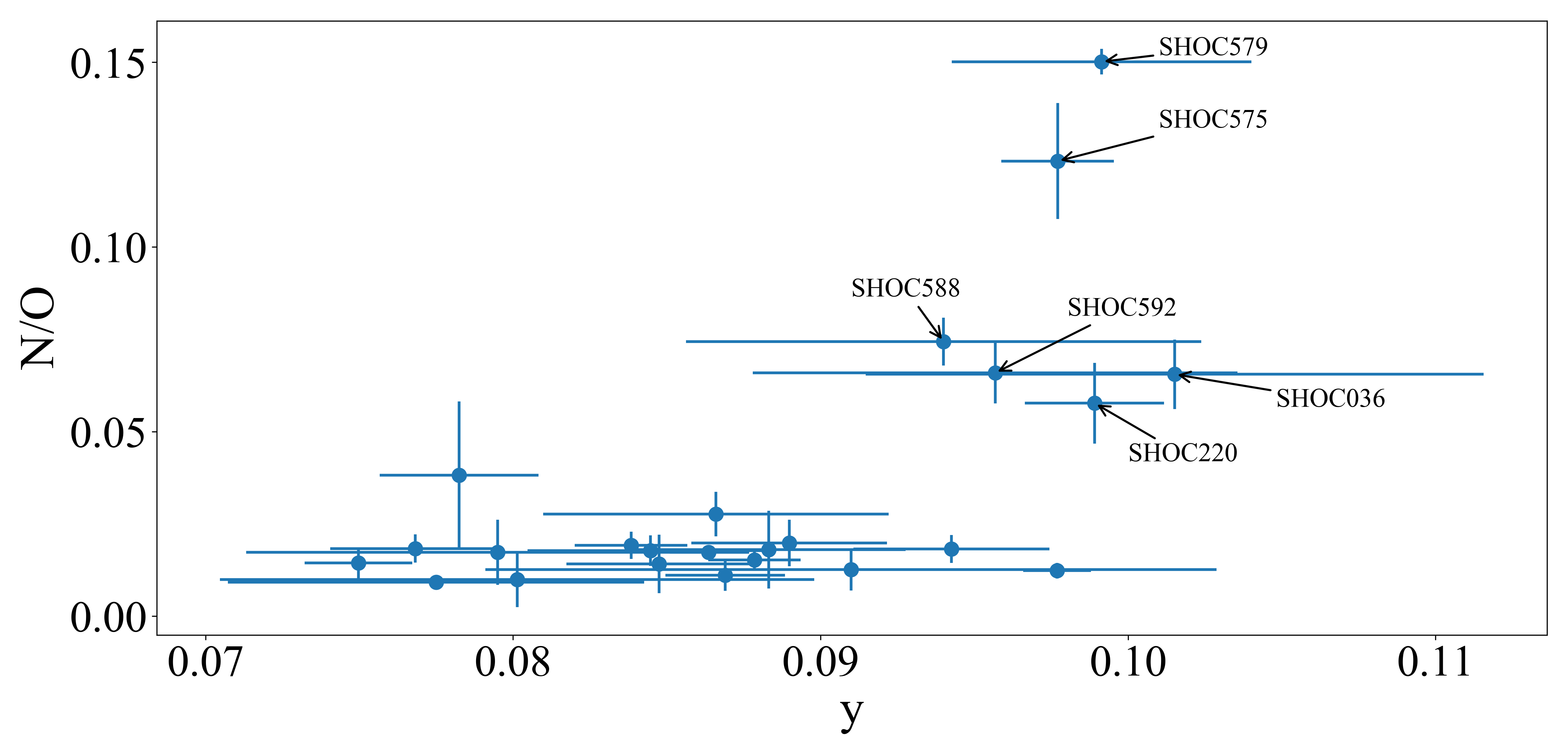}
\par\end{centering}
\caption{\label{fig:Nitrogen-to-oxygen_vs_He}Nitrogen to oxygen ratio versus
helium abundance for our sample. The labeled objects have been excluded from the linear regressions as outliers (see text).}
\end{figure*}
 In all three panels in Fig. \ref{fig:Primordial-linear-regressions}
a few objects appear to show a helium abundance distinctively above the linear fit.
These Y-enhanced objects occupy a different position for each regression:.
They fall in the low metallicity
edge of the  
$He-O$ plot bue on the high metallicity one for the $He-N$ plot. Fig. \ref{fig:Nitrogen-to-oxygen_vs_He}
shows the $\nicefrac{N}{O}$ ratio versus $y$. The majority of the
objects occupy a narrow band, $\Delta\nicefrac{N}{O}\approx0.025$ of about but
for those objects with $y>0.09$, the $\nicefrac{N}{O}$ ratio has a larger dispersion. This phenomenon is observed in both the low and high metallicity galaxies. Three of these objects:
$SHOC575$, $SHOC579$ and $SHOC022$, are identified as Wolf-Rayet galaxies ($WR038$, $WR039$ and $WR057$) in \cite{brinchmann2008}. 
Detecting WR features using  single slit observations is a challenging
task, since the slit must be precisely located in order to find 
the stars. Even the well observed galaxy $IZw18$ defied WR detection for a long time, until the work by \citet{izotov1997} and \citet{legrand1997}.
These three WR galaxies are included in \cite{perez-montero2013-a} in 
their study of the pollution of the interstellar medium by WR stars.
Our derived values $log\left(N/O\right)=-0.91\pm0.05$, $-0.82\pm0.1$,
$-1.71\pm0.08$ agree within uncertainties with theirs $log\left(N/O\right)=0.87\pm0.27$,
$-0.80\pm0.20$, $-1.48\pm0.26$ obtained on the basis of integral
field spectroscopy. \cite{perez-montero2013-a} concluded that while
the oxygen abundance in these galaxies is uniform in scales of the
order of several kpc, the $N/O$ ratio is not homogeneous. Higher
nitrogen excess was found at the spaxels close to the WR. Further
study, beyond the scope of this work, will probably determine how
these objects attained their helium enrichment, but it is clear that
these objects have an impact on the regressions. On the $Y-O$ fit
these galaxies tend to increase the $Y_{P}$ extrapolation while
on the $Y-N$ relation they decrease it. 
Consequently, we have excluded the six galaxies
with N/O excess (see Fig. \ref{fig:Nitrogen-to-oxygen_vs_He}) from all the $Y_{P}$ linear regressions.
These are: SHOC579, SHOC575, SHOC588, SHOC592, SHOC036 and SHOC220. SHOC022 and IZwicky18 are still included even when they show the presence of
WR stars. As they don't present any anomaly in their $N/O$ values, we
assumed that our observations do not include contaminated regions.
In the $Y-S$ regression these objects seem to have a helium enrichment. In contrast, their sulfur abundances do not seem to diverge from the rest of the sample, unlike in the oxygen and nitrogen cases. For consistency, however, these objects
were excluded as outliers from all the linear regressions. Still, they have
been included in Fig. \ref{fig:Primordial-linear-regressions} with dashed uncertainty bars.

\subsection{Linear regression}

The helium abundance, helium mass fraction and the abundances of O, N and S are given in 
Table \ref{tab:Elemental-abundances-and-Y}.
Two helium mass fractions were calculated for each object. From the
mass fraction formula $\left(X+Y+Z=1\right)$ it can be derived that:
\begin{equation}
Y=\frac{4\frac{HeI}{HI}\left(1-Z\right)}{1+4\frac{HeI}{HI}}\label{eq:Y_calculation}
\end{equation}
where it is common to express Z as a function of the oxygen abundance:
$Z\thicksim20\nicefrac{O}{H}$ or a similar expression using N
(see eq. \ref{eq:NO_relation}). \cite{izotov2013}
suggest a modification of the classical expression using an empirical
oxygen-metallicity relation, $Z=f\left(O/Z\right)$. In order to use
sulphur as an alternative metallicity tracer we define: 
\begin{equation}
Z\sim20\frac{O}{S}\cdot\frac{S}{H}\label{eq:OS_relation-1}
\end{equation}

The oxygen to sulphur ratio was studied by \cite{diaz1991} for the
high metallicity star forming regions in $M51$. Their result was
confirmed recently by \cite{dorsjr.2016} for a large sample of
Blue Compact Dwarfs. They obtain $\left\langle log\left(S/O\right)\right\rangle =-1.53\pm0.03$
and $\left\langle log\left(S/O\right)\right\rangle =-1.78\pm0.02$
for high and low metallicity HII regions respectively. The low metallicity
value is used in eq. \ref{eq:OS_relation-1} for our objects. The 
Y values calculated using O or S (see Table \ref{tab:Y_table}) differ
on a slightly larger dispersion when using S, where the error
on the $S/O$ gradient is taken into account. To
compute the linear regression we used the $curvefit$ routine in the
scipy package \citep{jones2001} within a bootstrap algorithm: in
a ten thousand iterations loop, a linear fit was performed on the
helium mass fractions and heavy element abundances, generated
from a normal distribution for each galaxy. This process was repeated
for each $Y-Z$ combination. From the output distributions, we determined
the mean $Y_{P,\,Z}$ and the $1\sigma$ uncertainty for each
element regression.

The first three rows in Table \ref{tab:Y_table} 
show the primordial helium abundance determinations using one of the three
elements as metallicity proxy. Column (3) in this table displays the number of objects
included in that regression. A good agreement
is obtained from the three regressions: $Y_{P,\,O}=0.246\pm0.005$,
$Y_{P,\,N}=0.251\pm0.005$, $Y_{P,\,S}=0.244\text{\ensuremath{\pm}}0.006$
within their uncertainty. By
using O and S as metallicity proxy to determine $Y_{P},$ we obtain
similar values, while using N results in a slightly larger $Y_{P}$.
Limiting the regressions to the 17 objects in our sample having oxygen, nitrogen and sulfur abundance determinations we get slightly
larger mean values: $Y_{P,O}=0.248\pm0.006,$ $Y_{P,\,N}=0.252\pm0.005$
and $Y_{P,\,S}=0.246\pm0.006$. As already discussed, most of the sample
for this work has been obtained with observations designed ad-hoc
to measure the {[}SII{]} and {[}SIII{]} lines, and as a consequence
our preferred regression is the one obtained using S:
\begin{equation}
Y_{P,\,S}=0.244\text{\ensuremath{\pm}}0.006
\end{equation}

We obtain good agreement within uncertainties with the most recent
work on the topic using a $Y-O$ regression. The latest result from
\cite{peimbert2016} is $Y_{P,\,O}=0.2446\text{\textpm}0.0029$
including the effect of temperature fluctuations (that would decrease  $Y_{P}$
by a value  between $0.003$ and $0.006$). The MCMC helium abundance determination 
from \cite{aver2015} gives $Y_{P,\,O}=0.2449\pm0.004$
using observations and the oxygen abundance calculation from 
\cite{izotov2014}. Izotov and colleagues  obtain 
$Y_{P,\,O}=0.2551\pm0.0022$, a higher value which
they claim it might require non-standard physics at the time of Big
Bang nucleosynthesis. Their regression includes the largest number 
of objects and their observations reached the near-IR $HeI\lambda10830\AA$
line. 
Assuming a standard
cold dark matter paradigm the Planck collaboration provides a (model
dependent) value of $Y_{P,\,BBN\,Planck}=0.24467\text{\textpm}0.0002$
\cite[see][]{planck-collaboration2015}. All three regressions agree with this result,
within their errors. The mean values obtained both using $S$ and
$O$ are closer to the Planck value, while we obtain a larger $Y_{P}$
value using $N$. This could be due to a selective N enrichment in
these objects, but chemical evolution corrections fall beyond our
analysis, on account of the uncertainties in all $Y_{P,Z}$ determinations.

It may be appreciated from this analysis, that deviant
objects both in Y and Z have a large impact in the regression. This
is partly due to the restricted range in metallicity for our sample
objects. It would be interesting to include in the future higher metallicity
objects (see Carigi \& Peimbert
2010) even when for such cases deriving the O abundance will become
more difficult \citep[see, e.g.][]{haegele2008}. One of the advantages of including
sulphur is that its abundance can be measured with reasonable accuracy
in metal rich, lower ionization HII regions, increasing the baseline
for the linear fit.

As discussed by \cite{peimbert2007}
the third most important source of error in determining $Y_{P}$ lies
in the extrapolation of the Y values to zero metallicity. To add further insight into this uncertainty
source, we propose to apply a multiple/multivariable linear regression
model. In this framework, a linear relation is established between
multiple independent variables and a dependent variable expressed
as a scalar. The approach is valid for this analysis as the abundances
involved are independent. In these fittings, we employ the
helium fraction by mass calculated using sulfur $\left(Y=Y_{S}\right)$
and we apply a similar boot-strap algorithm to estimate $Y_{P}$.
The possible combinations are displayed in Table \ref{tab:Y_table}. Even though
including additional elements does not seem to reduce the uncertainty,
the predicted $Y_{P}$ agrees better with the results by \cite{aver2015}
and \cite{peimbert2016}. This is particularly
true for the $Y_{P,\,O-N-S}=0.245\text{\ensuremath{\pm}}0.007$
regression. This result vindicates the addition of more elements as
metallicity tracers, in particular sulphur. In a forthcoming paper we plan to apply MCMC
sampling to the $Y-Z$ regressions in order
to explore how the addition of more elements affects the $Y_{P}$
calculation.

\begin{table}
\caption{\label{tab:Y_table}Primordial helium abundance determinations from
 linear regression combinations and comparison with the literature.}
\begin{centering}
{\begin{tabu}{lcc}%
\hline%
Elemental regression&Magnitude&Number of objects\\%
\hline%
$Y_{P,\,O}$&$0.246\pm0.005$&18\\%
\hline%
$Y_{P,\,N}$&$0.251\pm0.005$&18\\%
\hline%
$Y_{P,\,S}$&$0.244\pm0.006$&21\\%
\hline%
$Y_{P,\,O-N}$&$0.247\pm0.006$&17\\%
\hline%
$Y_{P,\,O-S}$&$0.244\pm0.006$&18\\%
\hline%
$Y_{P,\,N-S}$&$0.250\pm0.007$&18\\%
\hline%
$Y_{P,\,O-N-S}$&$0.245\pm0.007$&17\\%
\hline%
\hline%
$Y_{P,\,O}^{1}$&$0.2446\pm0.0029$&5\\%
\hline%
$Y_{P,\,O}^{2}$&$0.2449\pm0.004$&15\\%
\hline%
$Y_{P,\,O}^{3}$&$0.2551\pm0.0022$&28\\%
\hline%
$Y_{P,\,Planck BBN}^{4}$&$0.24467\pm0.0002$&-\\%
\hline%
\end{tabu}}
\par\end{centering}
[1] \cite{peimbert2016} [2] \cite{aver2015}
[3] \cite{izotov2014} [4] \cite{planck-collaboration2015}
(This value represents an upper limit from the four $\Lambda CDM$
parameter configurations presented by the authors)
\end{table}

\section{Conclusions}

We observed with ISIS in the WHT in two different configurations, 27 low metallicity young  HII
galaxies selected from SDSS - DR12. The objective was to reach the
near-IR $\left[SIII\right]\lambda\lambda9069,9532\AA$ lines to
estimate $T\left[SIII\right]$, the abundances of S and He and to
derive $Y_{P}$, the primordial abundance of helium from $\text{\ensuremath{\nicefrac{dY}{dZ}}}$
using for the first time $S$ as a tracer of metal abundance.

The selected sample with $\left(EW\left(H\alpha\right)>200\AA\right)$
represents very young bursts. In these objects, the nebular continuum
contributes as much as the youngest stellar population to the continuum
and needs to be taken into account.

The nebular continuum was calculated from first principles and removed
from the observed spectra. The helium abundance was derived using
the direct method. The underlying absorption on the recombination
lines was subtracted using a SSP fit which provided a complete characterization
of the stellar population for the spectra obtained with the 
instrumental setup that covered a wide spectral base. For the configuration
that covered a smaller wavelength range, the older stellar population
was more difficult to characterize and subtract.
We hope this will improve in the near future with stellar bases which go
beyond the $7000\AA$ boundary.

We could determine from our data two temperature diagnostics: $T\left[OIII\right]$
and $T\left[SIII\right]$. A model with two ionization zones was considered.
When the errors for both temperatures were significantly different,
an empirical relation was used to derive the temperature from the
more precise determination and the error was propagated.

We determined the ionic and total abundances for oxygen and nitrogen.
For sulphur, in high ionization zones the $S^{3+}$ component cannot
be neglected. We proposed an $ICF\left(S^{3+}\right)$ based on the
$Ar^{3+}/Ar^{2+}$ ratio, appropriate for our wavelength range. This
was derived using Cloudy photo-ionization models \citep{ferland2013}
with PopStar evolutionary models  \cite{martin-manjon2010}.

Linear regressions were found between Y and O, Y and N and Y and S.
Six objects: SHOC036, SHOC220, SHOC592, SHOC588, SHOC575 and SHOC579
show helium excess. Three of them are known to have WR stars
probably contaminating the interstellar medium. These objects also
show a nitrogen to oxygen excess $\left(N/O>0.04\right)$, and $\left(\nicefrac{He}{H}>0.09\right)$.
They were excluded from the seven linear regressions.

The results for primordial helium from the three regressions are $Y_{P,\,O}=0.246\pm0.005$,
$Y_{P,\,N}=0.252\pm0.005$, $Y_{P,\,S}=0.244\text{\ensuremath{\pm}}0.006$.;
they agree with each other within errors and also with the derived
value from Planck within standard Big Bang Nucleosynthesis  prescriptions. A multivariable regression
using all three elements results in $Y_{O-N-S}=0.245\text{\textpm}0.007$.
The novel use of S as proxy for the metal abundance and to derive
$Y_{P}$ has proved successful.

In a forthcoming paper (Fern\'andez et al. in preparation) we will include
true error propagation between all the steps involved to correct
better the telluric features and to improve the sulphur abundance
determination. We are also improving the fit for the underlying stellar
population and follow \cite{izotov2014} and \cite{aver2015} multidimensional
schemes to compute the helium abundance. This procedure can simultaneously
quantify the different processes contributing to the emission in the helium
lines.

As future improvements to the programe, we plan to include in forthcoming
observation campaigns, the relatively strong $HeI\lambda10830\mathring{A}$
line with high sensitivity to electron density and obvious advantage
to derive helium abundance \citep[see][]{izotov2014,aver2015}. We will also use new stellar libraries that extend up to near-infrared wavelengths to subtract better the stellar population contamination to the emission line spectrum.
We are aware of tools such as skycorr developed by \citet{noll2014},
which can model and remove airglow contamination. This contribution
is instrument-independent and it will allow in the future to propagate
the error in the sky subtraction procedure. 

\section*{Acknowledgments}

We are thankful to Enrique P\'erez-Montero and Daniel Miralles for their generous
discussions on the nebular and stellar continua determination and to Manuel Emilio Moreno Raya for his company during the first observation on the William Herschel telescope.

The authors also wish to thank an anonymous referee whose comments helped to improve the clarity of the paper.

We thank the Spanish
allocation committee (CAT) for awarding observing time and the cheerful technical support from the observatory personnel. 
Vital Fern\'andez is grateful to the Mexican research Council (CONACYT)
for suporting this research through studenship 554031/300844. Elena Terlevich and Roberto Terlevich acknowledge CONACYT for supportig this research under grants: CB-2008-103365. This work has been supported by DGICYT grants AYA2013-47742-C4-3-P and AYA2016-79724-C4-1-P. Partial financial support came also from proyect SELGIFS: PIRSES-GA-2013-612701-SELGIFS. Vital Fern\'andez is grateful to the hospitality of the Departamento
de F\'\i sica Te\'orica at the Universidad Aut\'onoma de  Madrid, Spain.

The WHT is operated in the island of La Palma by the Isaac Newton
Group in the Spanish Observatorio del Roque de los Muchachos of the
Instituto de Astrof\'{\i}sica de Canarias.

Funding for the creation and distribution of the SDSS Archive has
been provided by the Alfred P. Sloan Foundation, the Participating
Institutions, the National Aeronautics and Space Administration,
the National Science Foundation, the US Department of Energy, the
Japanese Monbukagakusho, and the Max Planck Society. The SDSS Web
site is \href{http://www.sdss.org}{http://www.sdss.org}. 

This research has made use of the NASA/IPAC Extragalactic Database
(NED) which is operated by the Jet Propulsion Laboratory California
Institute of Technology, under contract with the National Aeronautics
and Space Administration.





\bibliographystyle{mn2e}
\bibliography{V_fernandez_Biblio.bib}

\label{lastpage}
\end{document}